# Anomalous size effects with fixed criticality in bistable flexible mechanical metamaterials

Zehuan Tang[a,b], Tingfeng Ma[a,*], Boyue Su[a], Qing Xia[c], Pengfei Kang[a], Bowei Wu[a]

[a] Zhejiang-Italy Joint Lab for Smart Materials and Advanced Structures, School of Mechanical Engineering and Mechanics, Ningbo University, Ningbo 315211, China

[b] Key Laboratory of Mechanics on Environment and Disaster in Western China, The Ministry of Education of China, Lanzhou University, Lanzhou 730000, China

[c] School of Civil Engineering and Transportation, South China University of Technology, Guangzhou, 510641, Guangdong, China

## ABSTRACT

When the structure deformation is dominated by the low-energy deformation mode, the structure hardens with the increase in the size (number of units) at small sizes. This anomalous behavior will eventually disappear with the decay length of the finite structure converging to a size-independent characteristic quantity, but the specific critical point at which the anomalous behavior disappears still cannot be accurately and concisely described. Here, under two steady states of the bistable chain, we observed anomalous size effects with constant and oscillating criticality (the proportion of inhomogeneous deformation), two criticalities exactly separate the increasing and decreasing intervals of stiffness variation. They are interrelated due to the implied symmetries between the two steady states. On the other hand, they are distinguished because of the opposite superposition modes under the two steady states. Specifically, the constant criticality corresponds to the anomalous size effect achieved by the competition mechanism, while the oscillating criticality reveals an anomalous size effect achieved by the new mechanism (cancellation mechanism). In the anomalous size effect achieved by the cancellation mechanism, the singular characteristics generated by the completely cancelled deformation make it very robust. This robustness reflects in that the anomalous effect is no longer limited to linear small deformation, but it can still be observed stably in nonlinear large deformation. Our study reinterprets the anomalous size effect at a quantitative level, and the proposed cancellation mechanism expands the possible application range of this anomalous effect.

## 1. Introduction

Flexible mechanical metamaterials (FlexMMs) usually refer to metamaterials composed of flexible hinges and stiffer elements (Deng et al., 2021), and current studies of such metamaterials have focused on porous structures, origami structures, and bistable plates. Taking the widely studied porous structure as an example, it has attracted attentions because of its negative Poisson ratio under uniaxial compression (Bertoldi et al., 2010). Subsequently, extensive researches have been carried out in porous structures, as well as in structures derived from them, such as adjustable Poisson ratios (Dudek et al., 2023, Khan et al., 2021), vector solitons (Deng et al., 2017, Li et al., 2021, Zhang et al., 2023), as well as recently sequential buckling shock-absorbing metamaterials (Liu et al., 2024) and modular shapeshifting metamaterials (Zhao et al., 2025).

In a series of studies on FlexMMs, flexible hinges are often modeled as a combination of multiple springs (Demiquel et al., 2024, Liang & Crosby, 2020, Jiao et al., 2024), where the parameters of the torsion spring are likely to have the greatest impact on the response (Meeussen et al., 2024, Mizzi & Spaggiari, 2020, Chen et al., 2022). When the torsional stiffness is zero (Zhou et al., 2021) or small enough to be negligible (Ma et al., 2018), the structure with geometric asymmetry will acquire non-trivial topological features, and then realize topologically protected zero-energy modes (Kane & Lubensky, 2014, Coulais et al., 2017) or solitons (Chen et al., 2014, Ma et al., 2023). However, except for purely mechanism materials connected by ideal hinges, it is difficult for the typical FlexMMs to meet the condition that the torsional stiffness is strictly zero (Meeussen et al., 2024). When the torsional stiffness is not negligible (Demiquel et al., 2023, Yi et al., 2023, Vahabi et al., 2016), usually in the static state, the structure can only support the low-energy deformation mode, which is a no propagating mode (Tang et al., 2024). Compared with the topological zero-energy mode, the low-energy deformation mode does not have the feature of localization on the specific side, but the property of decay only from the excitation point also makes it have a series of interesting mechanical behaviors. For example, low-energy deformation on both sides can form a domain wall that separates two phases, and the configuration of the domain wall can be regulated by placing defects at different locations (Deng et al., 2020). At the same time, the decay property also makes the low-energy deformation conformal, and its distribution is affected by the load distribution at the boundary (Czajkowski et al., 2022). Furthermore, nonlinearity also gives more possibilities for the distribution of low-energy deformation (Zhang et al., 2019).

Overall, due to low-energy deformation sensitive to defects or constraints (Florijn et al., 2014), boundary conditions (Zheng et al., 2022, Liang & Crosby, 2020) and nonlinearity (Deng et al., 2020), current researches on low-energy deformation mostly focus on achieving specific deformation distributions by using these properties, and then achieving different effective stiffness through different deformation distributions (Deng et al., 2020, Li et al., 2024). However, since the unit cells of mechanical metamaterials cannot be considered as vanishingly small like in ordinary crystals, the size of the structure has a non-negligible influence on the effective stiffness. For example, in small-size structures, the ordinary Cauchy continuum mechanics theory is no longer applicable (Kadic et al., 2018), and the effective properties (such as effective stiffness) of a small-size structure may be significantly different from those of a large size (Ziemke et al., 2019). In the monostable chain bearing low-energy deformation, the relationship between the stiffness and the structure size was further revealed (Coulais et al., 2018), the anomalous phenomenon that the structure stiffness increases with the size increase was captured by the discrete model, and an size-independent characteristic length scale (Frenzel et al., 2017) was defined in the continuum model to estimate the range of this anomalous phenomenon. Here, we also observed the anomalous size effect (the stiffness increases first and then decreases with the increase in size) in the bistable chain, in particular, under any initial configuration and any parametric flexural hinge, this behavior of stiffness increasing with size in the bistable chain has a fixed constant criticality or oscillating criticality. In further analysis, we found that the oscillating criticality actually corresponds to the compliance of the structure tending to zero, this unique property not only intuitively proves that the constant in the constant criticality can only take 0.5, but more importantly, it reveals a new mechanism (cancellation mechanism) to achieve the anomalous size effect. The anomalous size effect achieved by the cancellation mechanism has an extremely large maximum stiffness value, which makes it robust in nonlinear large deformation.

## 2. Bistable flexible mechanical metamaterials

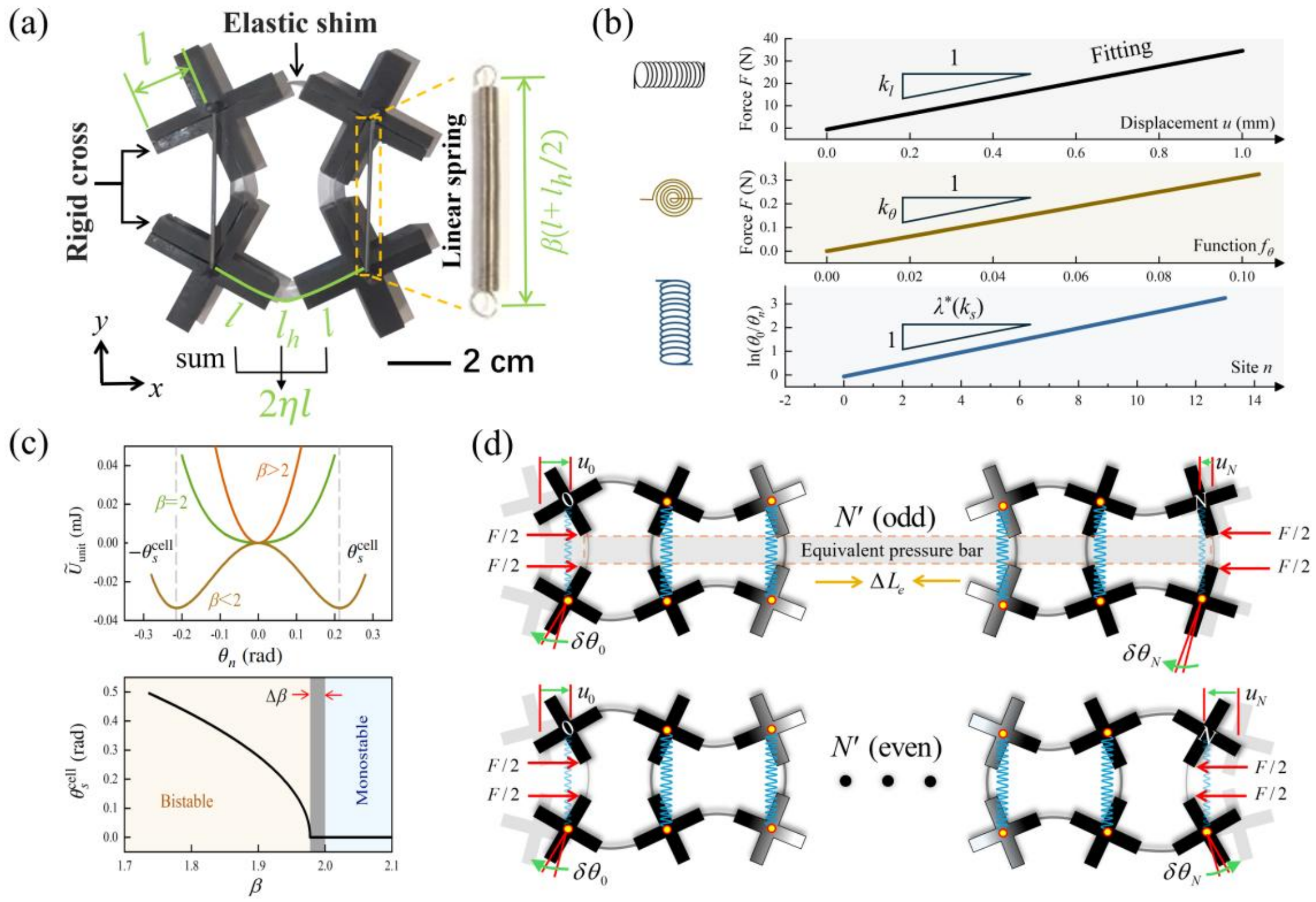

**Fig. (1).** (a) Bistable unit. (b) Three equivalent stiffness of shims, the experimental scheme refers to article (Tang et al., 2024). (c) The top sub-figure shows the elastic energy of a bistable unit $\widetilde{U}_{\text{unit}}$, where $\widetilde{U}_{\text{unit}} = 8k_\theta\theta_n^2 + k_t(\eta l)^2\left\{(2\cos\theta_n - \beta)^2 - (2-\beta)^2\right\}$. The bottom sub-figure shows the change of equilibrium angle with $\beta$. (d) Modeling schematic diagram of the odd-node chain and even-node chain.

The bistable units are shown in Fig. 1(a), which are assembled by rigid crosses and elastic shims. The arm length of the rigid cross is $l$, and the length and thickness of the shim are $l_h = 8$ mm and $l_t = 0.2$ mm, respectively. Fig. 1(a) also shows that after the cross and shim are assembled, the actual length of the moment arm of the cross is $\eta l$, where $\eta = 1 + l_h / (2l)$. The bistable property of units is realized by a pre-stretched linear spring connected between the center of the upper and lower crosses (Movie S1). The stiffness of the linear spring between the centers is $k_t =$ 0.15 N/mm, and its original length is $l_{h0}$, where $l_{h0}$ is expressed as $l_{h0} = \beta(l + l_h/2)$ by $\beta$. The shim is equivalent to a combination of stretching, torsional, and shearing linear springs, as shown in Fig. 1(b), and the equivalent stiffnesses of the three springs are $k_l = 35.14$ N/mm, $k_\theta = 3.17$ N·mm, and $k_s = 0.71$ N/mm respectively. We choose springs with different original lengths (namely different $\beta$) to achieve different number of steady states and equilibrium angles. As shown in the top sub-figure of Fig. 1(c), when $\beta \geq 2$ (the spring is pre-compressed), the unit has only one steady state $\theta_n = 0$. When $\beta < 2$ (the spring is pre-stretched), the unit has two symmetrical steady states $\theta_n = \theta_s^{\text{cell}}$ and $\theta_n = -\theta_s^{\text{cell}}$ ($\theta_s^{\text{cell}} > 0$), where the change of $\theta_s^{\text{cell}}$ with $\beta$ is shown in the bottom sub-figure of Fig. 1(c). The gray area with a width $\Delta\beta$ in the bottom sub-figure of Fig. 1(c) is the ineffective adjustment area of $\beta$, which appears because the moment arm ($\eta l \sin\theta_n$) of the pre-stretched force and the torque of the shim ($k_\theta \theta_n$) are same-order infinitesimal, and the pre-stretched force is not enough to make the cross rotate.

The bistable units are connected into a chain with $N'$ ($N' = N + 1$) nodes, as shown in Fig. 1(d). In order to ensure that each node has an equal equilibrium angle under the initial state, at both ends of the chain, the thickness of the shim connecting the upper and lower crosses is chosen to be $l_t / 2$ (the torsional stiffness of the shim with thickness $l_t / 2$ is equivalent to $k_\theta / 2$, in addition, changes in $k_l$ and $k_s$ of the shim at this position do not affect the response of the chain), while the stiffness of the linear spring between the centers is selected as $k_t / 2$, as shown by the light gray shim and light blue spring at both ends in Fig. 1(d). In Movie S2, the above design scheme is proved to be feasible by experiments and simulations. The equilibrium angles $\overline{\theta}$ of each node satisfy

$$6k_\theta\overline{\theta} = k_t(\eta l)^2\left(2\cos\overline{\theta} - \beta\right)\sin\overline{\theta}. \tag{1}$$

According to Eq. (1), the chain for $\beta < 2$ has two kinds of initial configuration, which are denoted as $\overline{\theta} = \theta_s$ and $\overline{\theta} = -\theta_s$.

A pair of forces $F$ are applied to both ends of the chain, as shown in Fig. 1(d), the equilibrium equations for the translation are established as follows (the derivation is detailed in Appendix A):

$$F/2 = -k_l\left[u_1 - u_0 + \eta l(\delta\theta_1 + \delta\theta_0)\sin\overline{\theta}\right], \tag{2a}$$

$$0 = -k_l\left[u_{n+1} + u_{n-1} - 2u_n + \eta l\left(\delta\theta_{n+1} - \delta\theta_{n-1}\right)\sin\overline{\theta}\right], n \in [1, N-1] \tag{2b}$$

$$-F/2 = k_l\left[u_N - u_{N-1} + \eta l(\delta\theta_{N-1} + \delta\theta_N)\sin\overline{\theta}\right]. \tag{2c}$$

The equilibrium equations for the rotation are as follows:

$$\begin{aligned} Fl\left(\eta\sin\overline{\theta} + \cos\overline{\theta}\right)/2 = {} & k_\theta(\delta\theta_1 + 2\delta\theta_0) - k_s l^2(\delta\theta_1 - \delta\theta_0)\cos^2\overline{\theta} \\ & -(k_t/2)(\eta l)^2\left[2\cos\left(2\overline{\theta}\right) - \beta\cos\overline{\theta}\right]\delta\theta_0, \end{aligned} \tag{3a}$$

$$\begin{aligned} F\eta l\sin\overline{\theta} = {} & k_\theta\left(\delta\theta_{n+1} + \delta\theta_{n-1} + 4\delta\theta_n\right) - k_s l^2\left(\delta\theta_{n+1} + \delta\theta_{n-1} - 2\delta\theta_n\right)\cos^2\overline{\theta} \\ & -k_t(\eta l)^2\left[2\cos\left(2\overline{\theta}\right) - \beta\cos\overline{\theta}\right]\delta\theta_n, n \in [1, N-1] \end{aligned} \tag{3b}$$

$$\begin{aligned} Fl\left(\eta\sin\overline{\theta} \pm \cos\overline{\theta}\right)/2 = {} & k_\theta(\delta\theta_{N-1} + 2\delta\theta_N) - k_s l^2(\delta\theta_{N-1} - \delta\theta_N)\cos^2\overline{\theta} \\ & -(k_t/2)(\eta l)^2\left[2\cos\left(2\overline{\theta}\right) - \beta\cos\overline{\theta}\right]\delta\theta_N. \end{aligned} \tag{3c}$$

## 3. Size effects under the two steady states

### *3.1. Numerical and experimental results*

Eqs. (2) and (3) are rewritten in the matrix form:

$$\begin{bmatrix} \mathbf{F} \\ \mathbf{M} \end{bmatrix} = \begin{bmatrix} \mathbf{K_r} & \mathbf{C} \\ \mathbf{O} & \mathbf{K_t} \end{bmatrix} \begin{bmatrix} \mathbf{U} \\ \boldsymbol{\delta\theta} \end{bmatrix}. \tag{4}$$

The one-dimensional chain in Fig. 1(d) is regarded as a pressure bar, and the length of the pressure bar is the distance between the acting points of forces at both ends. Under the action of force $F$, the change in bar length is denoted as $\Delta L_e$, and the effective stiffness of the one-dimensional chain can be written as

$$k_e = \frac{F}{\Delta L_e} = \frac{F}{u_0 - u_N + l(\delta\theta_0 \pm \delta\theta_N)\cos\overline{\theta}}. \tag{5}$$

Substituting the numerical solution of Eq. (4) into Eq. (5) yields the results of the discrete model.

For the odd-node chain, Fig. 2(b) shows the variation of its effective stiffness with size. Experimental and numerical results show that the effective stiffnesses under the two initial configurations are completely coincident, and they both decrease with the increase of size. This coincidence behavior is essentially due to the mirror symmetry between the two steady states of the odd-node chain, the schematic diagram of the mirror symmetry is shown in Fig. 2(a). The mirror symmetry makes the deformation distributions under $\overline{\theta} = \theta_s$ (denoted $u_n^+$ and $\delta\theta_n^+$) and the deformation distributions under $\overline{\theta} = -\theta_s$ (denoted $u_n^-$ and $\delta\theta_n^-$) satisfy the relationship $u_{N-n}^+ =$ $-u_n^-$ and $\delta\theta_{N-n}^+ = -\delta\theta_n^-$, and the above relation can also be obtained by the transformation $\bar{\theta} \to -\bar{\theta}$ in Eqs. (2) and (3). Fig. 2(c) shows this deformational symmetry between two steady states by using the angular deformation distribution as an example. In conclusion, the mirror symmetry of the odd-node chain ensures that the stiffnesses under the two initial configurations are equal.

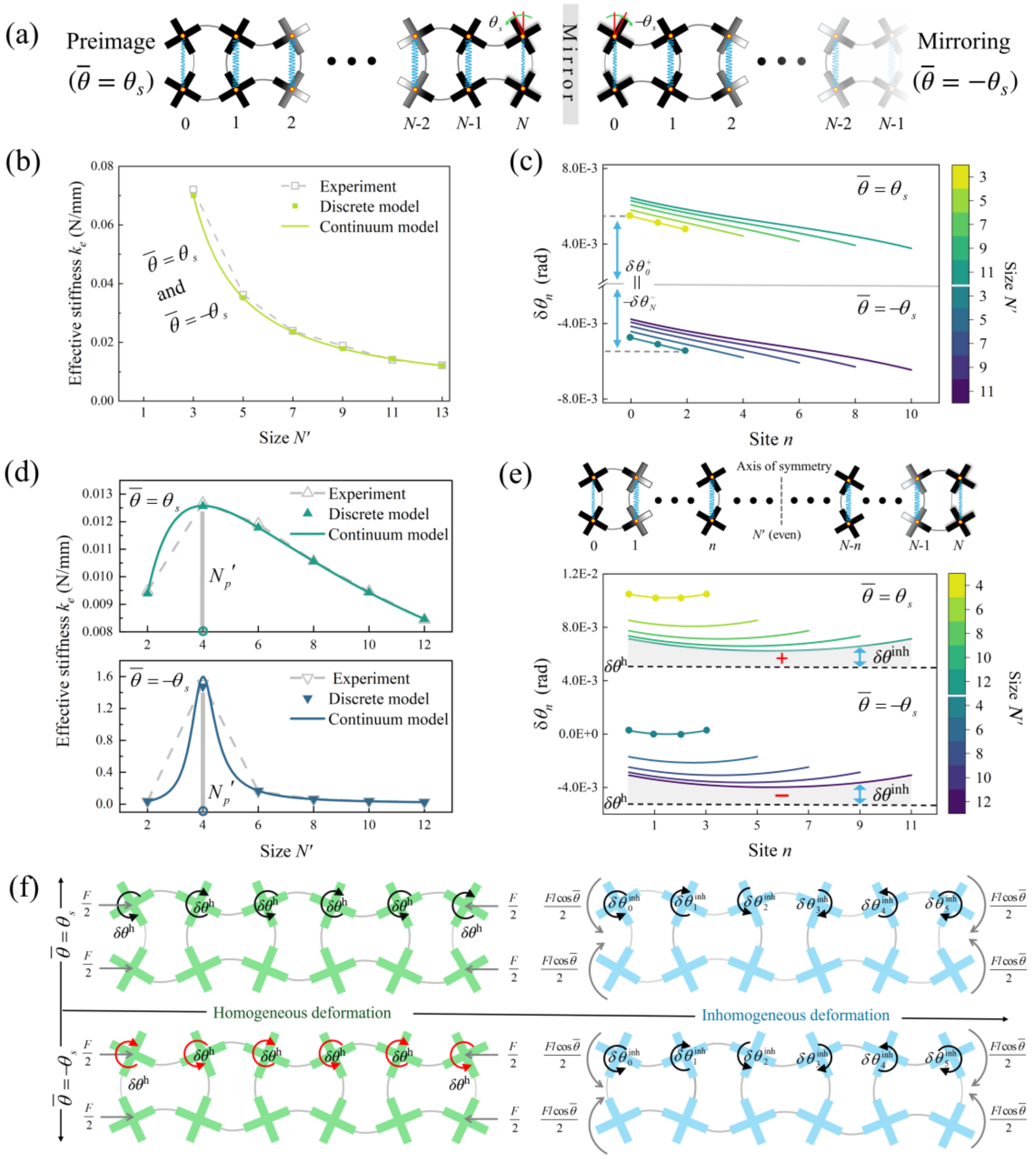

**Fig. 2.** (a) The odd-node chain has mirror symmetry between the two steady states. (b) In the odd-node chain, the effective stiffness varies monotonically with size, and the effective stiffnesses of two steady states are coincident. (c) The angle distributions under the two steady states have the relationship $\delta\theta_{N-n}^+ = -\delta\theta_n^-$. The solid lines in the figure represent the results of the continuum model (the continuum model will be discussed in the next section); the solid dots are the results of the discrete model, where the discrete points with $N' = 3$ are highlighted, while the other cases

($N' \neq 3$) are overlapped into the solid line because the results of the discrete and continuum models are highly consistent. (d) In the even-node chain, the effective stiffness varies non-monotonically with size, and the effective stiffnesses of two steady states do not coincide, but they have the same maximum point $N_P' = 4$. (e) The even-node chains are axisymmetric (top figure). The angle distribution has the relationship $\delta\theta_n^+ = \delta\theta_{N-n}^+$ (bottom figure), and the discrete points with $N' = 4$ in the figure are highlighted. The rotation deformation is regarded as a superposition of homogeneous deformation ($\delta\theta^{\mathrm{h}}$, dotted line in the figure) and inhomogeneous deformation ($\delta\theta^{\mathrm{inh}}$, light gray fill), where the values of $\delta\theta^{\mathrm{h}}$ is obtained through discrete models. Under the steady state $\bar{\theta} = \theta_s$ ($\bar{\theta} = -\theta_s$), $\delta\theta^{\mathrm{h}}$ and $\delta\theta^{\mathrm{inh}}$ produce positive (negative) superposition, they are represented in the figure by the marks "+" and "−", respectively. (f) The eccentric excitation is equivalent to the combination of the concentrated force $F/2$ through the center and the moment $(Fl\cos\bar{\theta})/2$, where the response of $F/2$ corresponds to the homogeneous deformation (green cross), and the response of $(Fl\cos\bar{\theta})/2$ corresponds to the inhomogeneous deformation (light blue cross). The two responses under steady state $\bar{\theta} = \theta_s$ and $\bar{\theta} = -\theta_s$ generate a superposition of type "+" and type "−", respectively.

Different from the above case of odd nodes, the effective stiffnesses of the even-node chain do not coincide under the initial configurations $\bar{\theta} = \theta_s$ and $\bar{\theta} = -\theta_s$, as shown in Fig. 2(d). The total deformation of the chain is resolved into homogeneous deformation and inhomogeneous deformation as shown in Fig. 2(f) (Movie S3 shows the homogeneous deformation of the chain, and Movie S4 shows the inhomogeneous deformation of the chain under unilateral moment excitation). This non-coincidence behavior comes from the different superposition modes of these two deformation components. Specifically, the homogeneous deformation and inhomogeneous deformation under $\bar{\theta} = \theta_s$ have the same contribution to $\Delta L_e$, as shown in Fig. 2(f). In this case, these two deformation components produce the positive superposition, which is represented by the mark "+" in Fig. 2(e). While under $\bar{\theta} = -\theta_s$, the contributions of the two deformation components to $\Delta L_e$ are opposite, as shown in Fig. 2(f), in which the two deformation components generate the negative superposition, as represented by the mark "−" in Fig. 2(e). As shown in Fig. 2(e), the different superposition modes lead to the difference between the total deformation distributions under the two initial configurations, resulting in the chain exhibiting different stiffness under two initial configurations.

In the case of even nodes, the effective stiffnesses of two steady states are not equal, but both exhibit anomalous size effects, and they have the same maximum point $N_P{}' = 4$, as shown by the gray solid line in Fig. 2(d). In other words, the chains under the two steady states have the same optimal size to achieve the maximum stiffness, this special phenomenon will be discussed in Section 3.2 based on the continuum model.

### *3.2. Continuum model for even nodes*

For the equilibrium equations Eqs. (2b) and (3b), the form in the continuum limit is considered,

$$\frac{\partial^2 u}{\partial X^2} = -2\eta l \sin\overline{\theta}\frac{\partial \delta\theta}{\partial X}, \tag{6}$$

$$\frac{\partial^2 \delta\theta}{\partial X^2} - \frac{1}{\left[\lambda^*(K_1, K_2)\right]^2}\delta\theta = -\frac{F\eta l \sin\overline{\theta}}{(K_2\,/\,2)}, \tag{7}$$

where, $K_1 = K_1(k_\theta, k_s, k_t)$ and $K_2 = K_2(k_\theta, k_s)$ are two stiffness parameters defined in $\mathbf{K_r}$ (the specific expressions are detailed in Appendix A). $\lambda^*(K_1, K_2)$ in Eq. (7) is the characteristic decay length of the inhomogeneous deformation, and

$$\lambda^*(K_1, K_2) = \sqrt{\frac{(K_2\,/\,2)}{K_1 - K_2}}. \tag{8}$$

In the structure deformation, $\lambda^*$ is the length required for the infinite chain to decay a unit deformation in logarithmic coordinates.

For the even-node chain, as shown in Fig. 2(e), it has axial symmetry, which simplifies the form of the angle solution, namely, $\delta\theta(X) = \delta\theta(N - X)$. The solution to Eq. (7) can be written as

$$\delta\theta = \delta\theta^{\mathrm{inh}} + \delta\theta^{\mathrm{h}} = C_1\left[e^{X/\lambda^*} + e^{(N-X)/\lambda^*}\right] + C_\theta. \tag{9}$$

As can be seen from Eq. (9), unlike the monostable chain whose rotation deformation is purely composed of inhomogeneous deformation (Coulais et al., 2018), rotation deformation in the bistable chain has two parts: inhomogeneous deformation $\delta\theta^{\mathrm{inh}} = C_1\left[e^{X/\lambda^*} + e^{(N-X)/\lambda^*}\right]$ and homogeneous deformation $\delta\theta^{\mathrm{h}} = C_\theta$, which correspond to the response under the moment and the concentrated force through the center in Fig. 2(f) respectively. The easily solvable $C_\theta$ is first identified, as the particular solution of Eq. (7), which can be written as

$$C_\theta = \frac{(\lambda^*)^2 F\eta l \sin\overline{\theta}}{(K_2\,/\,2)} = \frac{F\eta l \sin\overline{\theta}}{K_1 - K_2}. \tag{10}$$

On the other hand, in a physical meaning, $C_\theta$ is the homogeneous rotation deformation of the chain after each node is subjected to the moment $F\eta l\sin\overline{\theta}$, which can also be directly determined by the equilibrium equation. Next, $C_1$ is determined by the equilibrium equations Eq. (3a) of the 0-th node, and

$$C_1 = \frac{Fl\cos\overline{\theta}}{K_1\left(1+e^{N/\lambda^*}\right) - K_2\left[e^{1/\lambda^*} + e^{(N-1)/\lambda^*}\right]}, \tag{11}$$

Substituting Eq. (9) into Eq. (6), the displacement solution can be obtained, namely,

$$u = u^{\text{inh}} + u^{\text{h}} + D_2 = -2\eta l\lambda^* C_1\left[e^{X/\lambda^*} - e^{(N-X)/\lambda^*}\right]\sin\overline{\theta} - D_1 X + D_2. \tag{12}$$

Similarly, unlike the monostable chain whose displacement deformation is purely described by homogeneous deformation, due to the coupling of translation and rotation, the displacement deformation in the bistable chain has inhomogeneous deformation part $u^{\text{inh}} = -2\eta l\lambda^* C_1\left[e^{X/\lambda^*} - e^{(N-X)/\lambda^*}\right]\sin\overline{\theta}$ and homogeneous deformation part $u^{\text{h}} = -D_1 X$.

In Eq. (12), $u^{\text{inh}}$ is generated by the reduction of the horizontal projection of the rigid cross. After each node of the chain rotates at the corresponding angle $\delta\theta^{\text{inh}}$ (namely, the $n$-th cross rotates at the angle $\delta\theta^{\text{inh}}(n)$, $n \in [0, N]$), the displacement difference between the 0-th node and the $N$-th node (denoted as $\Delta L_0$) is

$$\Delta L_0 = u^{\text{inh}}(0) - u^{\text{inh}}(N) = \left(\eta l\sin\overline{\theta}\right)\sum_{0}^{N-1}\left[\delta\theta^{\text{inh}}(n) + \delta\theta^{\text{inh}}(n+1)\right], \tag{13}$$

where, $\sum_0^{N-1}\left[\delta\theta^{\text{inh}}(n) + \delta\theta^{\text{inh}}(n+1)\right]$ can be obtained by summing Eqs. (3) (see (i) in Appendix B for details), $\Delta L_0$ can be further expressed as

$$\Delta L_0 = 2lC_\theta\cos\overline{\theta}. \tag{14}$$

The result of Eq.(14) shows that the value of $u^{\text{inh}}(0) - u^{\text{inh}}(N)$ is independent of the size $N'$. Namely, when the initial configuration of the chain is fixed, no matter how the number of nodes changes, the end-to-end displacement difference generated by the inhomogeneous rotation deformation under the fixed excitations is always a constant. This is the most basic anomalous property to the bistable chain. In Movie S5, this anomalous property is verified by simulations.

For the homogeneous deformation part $u^{\text{h}}$ in Eq. (12), the expression of the integral constant $D_1$ is

$$D_1 = 2\eta lC_\theta\sin\overline{\theta} + F/(2k_l), \tag{15}$$

where $2\eta l C_\theta \sin\overline{\theta}$ is the reduction of the horizontal projection after single cross is rotated by angle $C_\theta$, $F/(2k_l)$ is the reduction in the length of a single shim under pressure $F/2$. $D_2$ corresponds to the rigid body translation mode generated by the rank-deficient matrix $\mathbf{K_t}$ (the structure lacks a horizontal constraint), which can take any value because it is eliminated in Eq. (5). In summary, the four integral constants in the continuum model are all determined. The complete continuum model provides a contrast to the numerical and experimental results, as shown in Fig. 2(d). Next, it will be used to analyze the maximum point of the stiffness.

In order to carry out the analysis of the maximum point more simply, the effective compliance of the chain with the initial configuration of $\overline{\theta}$ and nodes of $N'$ is defined as $\eta_e\left(\overline{\theta}, N'\right) = \left[k_e\left(\overline{\theta}, N'\right)\right]^{-1}$ (see (i) in Appendix B for details),

$$\eta_e\left(\overline{\theta}, N'\right) = \frac{1}{F}\left[2\Delta L_0\left(\overline{\theta}\right) + 2l\delta\theta_0^{\mathrm{inh}}\left(\overline{\theta}, N'\right) + D_1\left(\overline{\theta}\right)N\right], \tag{16}$$

where, the definition of $\delta\theta_0^{\mathrm{inh}}$ in Eq. (16) is $\delta\theta_0^{\mathrm{inh}} = \delta\theta^{\mathrm{inh}}(X)\,|_{X=0}$,

$$\delta\theta_0^{\mathrm{inh}}\left(\overline{\theta}, N'\right) = C_1\left(1 + e^{N/\lambda^*}\right). \tag{17}$$

$\eta_e\left(\overline{\theta}, N'\right)$ is the change of the equivalent pressure bar under a pair of unit forces. In its expression Eq. (16), both $\delta\theta_0^{\mathrm{inh}}\left(\overline{\theta}\right)$ and $D_1\left(\overline{\theta}\right)$ have even symmetry. Although $\Delta L_0\left(\overline{\theta}\right)$ does not have even symmetry, it is size-independent, which makes the $\eta_e$ of the even-node chain have a special property, that is, $\partial_{N'}\left[\eta_e\left(\overline{\theta}, N'\right)\right] = \partial_{N'}\left[\eta_e\left(-\overline{\theta}, N'\right)\right]$. This property guarantees that the stiffness curves under $\overline{\theta} = \theta_s$ and $\overline{\theta} = -\theta_s$ have the same maximum point. Namely, the even-node chains under two stable states reach maximum stiffness at the same size, which explains that the two stiffness curves in Fig. 2(d) have the same maximum point.

In this section, the size independence of $\Delta L_0$ ensures that the compliances of two steady states have the same change rate, however, the special properties based on $\Delta L_0 = 2lC_\theta \cos\overline{\theta}$ in the bistable chain are not limited to this. In $\Delta L_e$, the inhomogeneous deformation part (denoted as $\Delta L_e^{\mathrm{inh}}$) and homogeneous deformation part (denoted as $\Delta L_e^{\mathrm{h}}$) are expressed as (see (ii) in Appendix B for details)

$$\Delta L_e^{\mathrm{inh}} = 2l\left(C_\theta + \delta\theta_0^{\mathrm{inh}}\right)\cos\overline{\theta}, \tag{18}$$

$$\Delta L_e^{\mathrm{h}} = D_1 N + 2lC_\theta \cos\overline{\theta}. \tag{19}$$

According to Eqs. (18) and (19), the anomalous size effect in Fig. 2(d) can still only be explained at the qualitative level. Namely, the homogeneous deformation always increases with the increase in size. Because $D_1 N$ in $\Delta L_e^{\mathrm{h}}$ can be equivalent to the response of the system comprised of $N$ springs in series, the greater number of springs in series, the greater the response deformation; While inhomogeneous deformation always decreases with the increase in size. This is because the small-size structures have a stronger superposition effect, which makes the small-size structures exhibit greater deformation by reducing the gradient of the deformation field (see (iii) in Appendix B for details). Competition between the two deformation modes results in anomalous size effects. But in Eqs. (18) and (19) we note two unusual results. The first result is that $\Delta L_e^{\mathrm{inh}}$ and $\Delta L_e^{\mathrm{h}}$ may change signs in the case of $\bar{\theta} = -\theta_s$. Taking $\Delta L_e^{\mathrm{inh}} = 2l\left(C_\theta + \delta\theta_0^{\mathrm{inh}}\right)\cos\bar{\theta}$ as an example, in the expression of $\Delta L_e^{\mathrm{inh}}$, the signs of $C_\theta$ and $\delta\theta_0^{\mathrm{inh}}$ are opposite under $\bar{\theta} = -\theta_s$ (as shown in the Fig. 2(f)), which provides a condition for the sign change of $\Delta L_e^{\mathrm{inh}}$. If their signs really change, then with the appearance of the zero deformation point, the above qualitative monotonicity analysis may fail in the stiffness analysis. The second result is that $\Delta L_0$ ($2lC_\theta \cos\bar{\theta}$ in both equations) makes a connection between inhomogeneous deformation and homogeneous deformation, which is different from the decoupling of inhomogeneous deformation and homogeneous deformation in the monostable chain. For the first result, quantitative analysis might be the only way to confirm whether monotonicity analysis is applicable. Coincidentally, in the second result, the connection established by $\Delta L_0$ exactly makes quantitative analysis for the anomalous size effect possible. So in following Section 4, we attempt to quantitatively reveal the mechanism of the anomalous size effects under the two steady states.

## 4. Criticality of stiffness transition

### *4.1. Constant criticality for $\bar{\theta} = \theta_s$*

Before discussing the mechanism of the anomalous size effect, it is necessary to determine its occurrence range. In Fig. 3(a), $k_e(2)$ and $k_e(4)$ are respectively the stiffnesses of the 2-node structure and of the 4-node structure. As shown in Fig. 3(a), if the stiffness increases first and then decreases in the discrete system (yellow line charts), this requires $k_e(2) < k_e(4)$ (as indicated by the upward brown arrow in the yellow line charts), which means that the boundary of the range in which anomalous size effect occurs can be given by $k_e(2) = k_e(4)$ (grey line charts). The results of the discrete model show that the boundary values under $\bar{\theta} < 0$ are opposite to those under $\bar{\theta} > 0$, and these two boundary values are denoted as $\bar{\theta} = -\bar{\theta}_{\text{an}}$ (where, $\bar{\theta}_{\text{an}} > 0$) and $\bar{\theta} = \bar{\theta}_{\text{an}}$ in Fig. 3(a). The variation curves of $\eta_e(\bar{\theta}, 2)$ and $\eta_e(\bar{\theta}, 4)$ with respect to $\bar{\theta}$ are shown in Fig. 3(b). Opposite boundary values limit the initial configuration with anomalous effects to a symmetric range, namely $[-\bar{\theta}_{\text{an}}, \bar{\theta}_{\text{an}}]$, and the intersection points of the two compliance curves determines $\bar{\theta}_{\text{an}} = 0.45$ (rad). The source of this symmetry is explained below. Although $\eta_e(\bar{\theta}, N')$ does not have any symmetry, in the compliance difference $\Delta\eta_e(\bar{\theta})$ ($\Delta\eta_e(\bar{\theta}) = \eta_e(\bar{\theta}, 2) - \eta_e(\bar{\theta}, 4)$) between the 2-node structure and the 4-node structure, the $\Delta L_0$ without even symmetry is eliminated in the compliance difference due to its size independence. Therefore, as shown in the inset of Fig. 3(b), $\Delta\eta_e(\bar{\theta})$ is even symmetric, which results in the range of the anomalous size effect being symmetrically distributed along the horizontal axis.

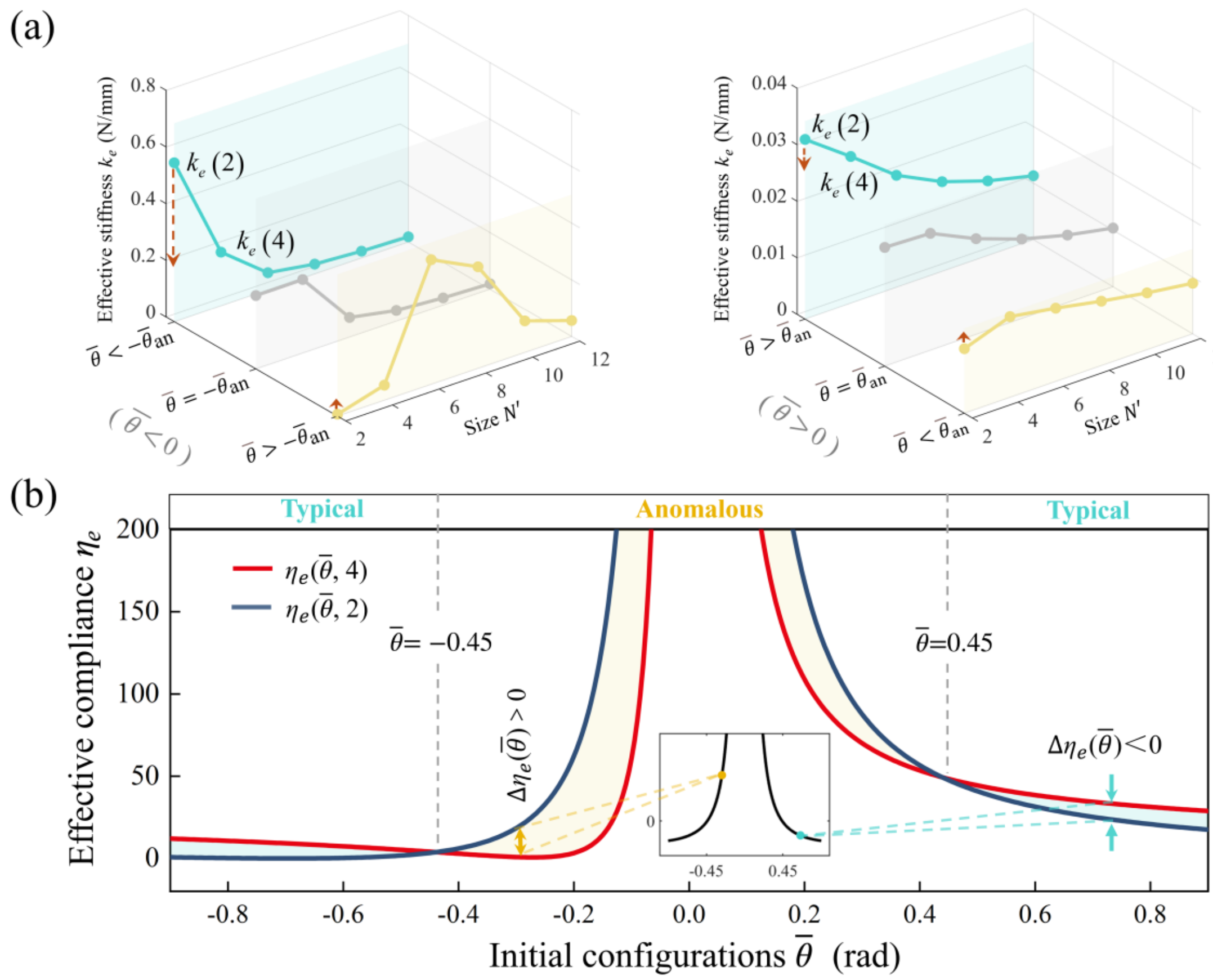

**Fig. 3.** (a) Size effects under different initial configurations, where the left sub-figure and the right

sub-figure represent the cases of $\overline{\theta} < 0$ and $\overline{\theta} > 0$, respectively. The solid points in the figure are the results of the discrete model. It can be determined whether $k_e(N')$ is monotonically variable by comparing the stiffness $k_e(2)$ of the 2-node structure with the stiffness $k_e(4)$ of the 4-node structure, as shown by the direction of the brown arrow in the figure. (b) The variation curves of compliance with the initial configuration, where the black-blue curve and the red curve represent compliances of the 2-node structure and the 4-node structure respectively, and these two curves intersect at $\overline{\theta} = 0.45$ and $\overline{\theta} = -0.45$. The inset shows the variation of the compliance difference $\Delta\eta_e(\overline{\theta})$ ($\Delta\eta_e(\overline{\theta}) = \eta_e(\overline{\theta}, 2) - \eta_e(\overline{\theta}, 4)$) with the initial configuration, and $\Delta\eta_e(\overline{\theta})$ is even symmetric.

Further, we discuss the criticality at which stiffness monotonicity transitions, and the case of $\overline{\theta} = \theta_s$ is discussed first. The inhomogeneous deformation serves as the only factor for stiffness increase, and an inhomogeneity index $P$ reflecting its proportion is defined,

$$P = \operatorname{sgn}\left(\frac{\Delta L_e^{\mathrm{inh}}}{\Delta L_e^{\mathrm{h}}}\right)\frac{|\Delta L_e^{\mathrm{inh}}|}{|\Delta L_e^{\mathrm{inh}}| + |\Delta L_e^{\mathrm{h}}|}. \tag{20}$$

$P \in [-1, 1]$, $|P| \to 1$ and $|P| \to 0$ correspond to inhomogeneous deformation and homogeneous deformation dominating in total deformation, respectively. In addition, $P > 0$ indicates that inhomogeneous deformation and homogeneous deformation produce positive superposition, and $P < 0$ indicates that they produce negative superposition. Fig. 4(a) shows the $P$-values under different initial configurations and sizes, the variation trend of index $P$ indicates that the proportion of inhomogeneous deformation decreases with the increase of initial angle and size. For such a $P$-values surface, the only possible certainty might be that $P = 1$ corresponds to stiffness increase and $P = 0$ corresponds to stiffness decrease, because they respectively correspond to inhomogeneous deformation and homogeneous deformation completely occupying the total deformation. Therefore, in order to explore the relationship between $P$-values and stiffness monotonicity more deeply, under each initial configuration, the derivative $(dk_e / dN')$ of stiffness with respect to size is plotted in Fig. 4(b), where the yellow (baby blue) region represents the increasing (decreasing) interval of stiffness. These two regions are separated by the white region between them, the white region is the transition criticality of the stiffness monotonicity. When the curve $P(\overline{\theta}, N') = 0.5$ in Fig. 4(a) is plotted to Fig. 4(b), we find that the curve is exactly located in the white region separating the increasing and decreasing interval, which indicates that the relationship between the proportion of inhomogeneous deformation and stiffness monotonicity can be quantified. Specifically, for the case of $\overline{\theta} < \overline{\theta}_{\text{an}}$ in Fig. 4(b) (the area below the grey dotted line), when the proportion of inhomogeneous deformation is higher than 0.5 (namely, at small sizes), the stiffness increases with the increase in size. If the size continues to increase, the proportion of inhomogeneous deformation will be lower than 0.5, and the stiffness will decrease with the increase in size. The above discussion reinterprets the anomalous size effect at a quantitative level and concludes that the stiffness monotonicity follows $P = 0.5$ as the transition criticality.

### *4.2. Oscillating criticality for $\overline{\theta} = -\theta_s$*

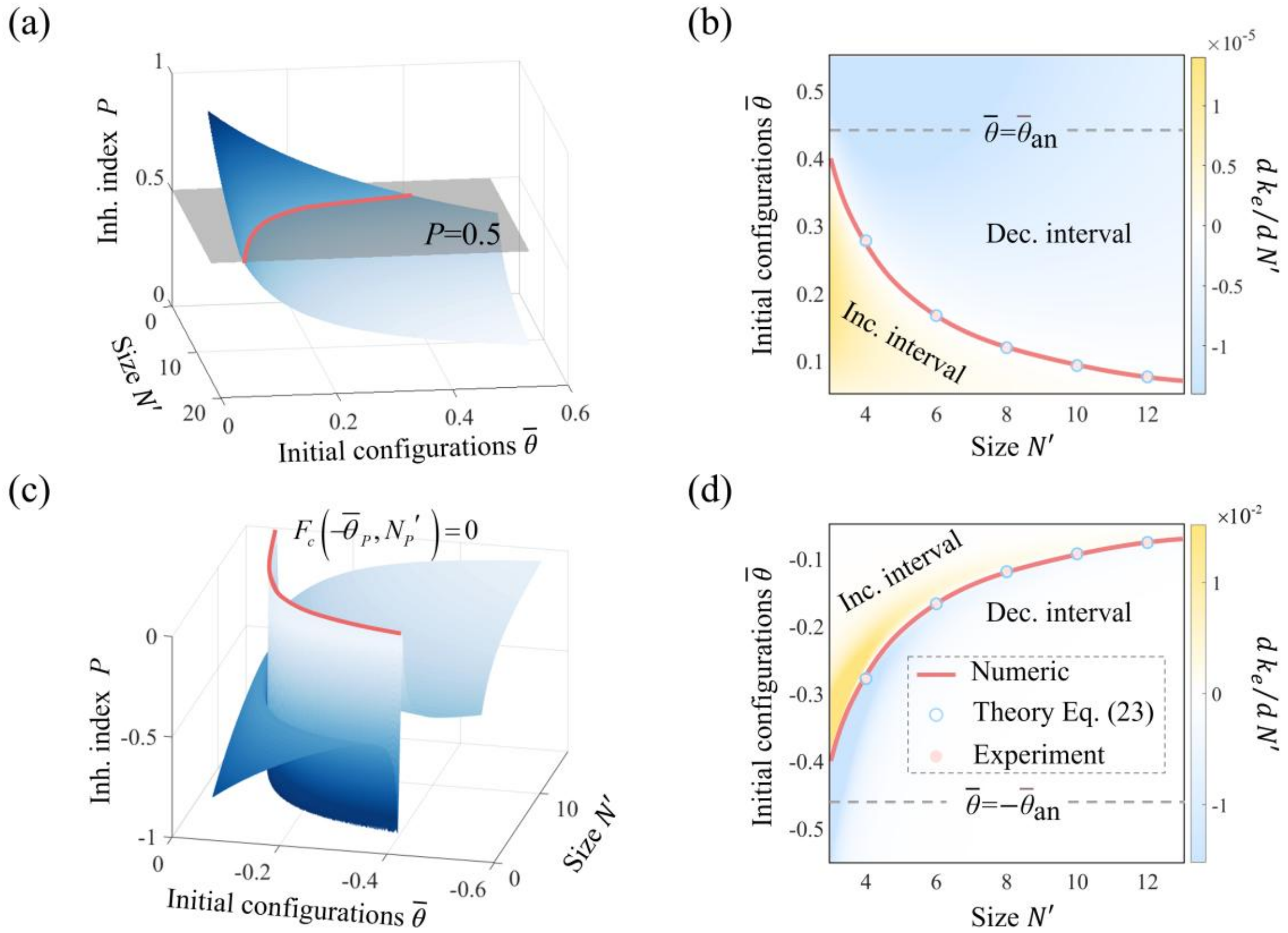

**Fig. 4.** (a) In the case of $\overline{\theta} = \theta_s$, the variation of inhomogeneity index $P$ with respect to initial configuration and size, the pink solid line in the figure is the curve $P(\overline{\theta}, N') = 0.5$. (b) The phase diagram of derivative $d[k_e(N')] \,/\, dN'$ in the case of $\overline{\theta} = \theta_s$, where the increasing interval and decreasing interval of $k_e(N')$ are separated by the curve $P(\overline{\theta}, N') = 0.5$. (c) In the case of $\overline{\theta} = -\theta_s$, the $P$-values surface is discontinuous, and the pink solid line $F_c(-\overline{\theta}_P, N_P') = 0$ is obtained by fitting the discontinuity points. (d) The derivative phase diagram in the case of $\overline{\theta} = -\theta_s$, the curve $F_c\left(-\overline{\theta}_P, N_P'\right) = 0$ separates the increasing and decreasing interval of stiffness. The azure hollow circle in the figure is the result of Eq. (23). Under the initial configuration corresponding to the azure hollow circle, the experimental results of the maximum point $N_P'$ are shown in the light pink dot in the figure.

As shown in Fig. 4(c), in the case of $\overline{\theta} = -\theta_s$, $P<0$, which indicates that there are cancellations between the inhomogeneous and homogeneous deformation. In Fig. 4(c), the $P$-values surface in this case is not continuous like the surface under $\overline{\theta} = \theta_s$, and the oscillating discontinuity point occurs as $P$ tends to $-0.5$. These discontinuity points are projected onto the $\overline{\theta} - N'$ plane and fitted into a continuous curve, as shown by the pink curve in Fig. 4(c). When this curve is drawn into the derivative phase diagram, it also exactly separates the increasing and decreasing interval of stiffness, as shown in Fig. 4(d). This indicates that the anomalous size effects under $\overline{\theta} = -\theta_s$ take the oscillating $P$-values as the criticality. Meanwhile, the oscillating $P$-values foreshadow that this anomalous effect might not be the product of the competition between two kinds of deformations. Because in the competition mechanism, conflicting indexes such as $P = -1$ (the stiffness must increase) and $P = 0$ (the stiffness must decrease) cannot occur at the same point. This singular phenomenon has caught our attention, and this phenomenon will be discussed in detail next.

The curve obtained by fitting is denoted as $F_c\left(-\overline{\theta}_P, N_P'\right) = 0$. According to the expression for $P$ (Eq. (20)), when $\left(\overline{\theta}, N'\right) \to \left(-\overline{\theta}_P, N_P'\right)$, the oscillating discontinuous point can only occur in the case where $\Delta L_e^{\mathrm{inh}}$ and $\Delta L_e^{\mathrm{h}}$ approach 0 simultaneously. In the $P$-values surface of Fig. 4(c), $\Delta L_e^{\mathrm{inh}}\left(\overline{\theta}, N'\right) = 0$ corresponds to the curve $P\left(\overline{\theta}, N'\right) = 0$, and $\Delta L_e^{\mathrm{h}}\left(\overline{\theta}, N'\right) = 0$ corresponds to the curve $P\left(\overline{\theta}, N'\right) = -1$. The almost coincident $P\left(\overline{\theta}, N'\right) = 0$ and $P\left(\overline{\theta}, N'\right) = -1$ in the $P$-values surface confirms it is possible for $\Delta L_e^{\mathrm{inh}}\left(\overline{\theta}, N'\right)$ and $\Delta L_e^{\mathrm{h}}\left(\overline{\theta}, N'\right)$ to approach 0 simultaneously, and their coincidence is necessarily accompanied by the oscillation of the $P$-values. This numerically indicates that the oscillation of $P$-values originates from $\Delta L_e^{\mathrm{inh}}$ and $\Delta L_e^{\mathrm{h}}$ tending to 0 simultaneously. However, whether it is universal for $\Delta L_e^{\mathrm{inh}}$ and $\Delta L_e^{\mathrm{h}}$ to tending to 0 at the same time needs further theoretical analysis.

According to Eq. (18), the condition for $\Delta L_e^{\text{inh}}=0$ is $\delta\theta_0^{\text{inh}}=-C_\theta$. Substituting Eqs. (10) and (17) into this condition (see Appendix C for details) yields

$$-\frac{1}{\eta\tan\overline{\theta}}=\frac{(K_1-K_2)\widetilde{\delta\theta}^{\text{inh}}(0)-K_2\left[\widetilde{\delta\theta}^{\text{inh}}(1)-\widetilde{\delta\theta}^{\text{inh}}(0)\right]}{(K_1-K_2)\widetilde{\delta\theta}^{\text{inh}}(0)},\tag{21}$$

where, $\widetilde{\delta\theta}^{\text{inh}}(X)=e^{X/\lambda^*}+e^{(N-X)/\lambda^*}$ is the normalized inhomogeneous angle distribution. Next, the condition for $\Delta L_e^{\text{h}}=0$ is discussed. Because the chain is connected through the low torsional shims, in $D_1$, $2\eta l C_\theta\sin\overline{\theta}\gg F/(2k_l)$. Namely, the tension deformation of the shim can be ignored (Tang et al., 2023). So $\Delta L_e^{\text{h}}$ is approximated as

$$\Delta L_e^{\text{h}}\approx 2lC_\theta\left(\eta N\sin\overline{\theta}+\cos\overline{\theta}\right),\tag{22}$$

the condition for $\Delta L_e^{\text{h}}=0$ can be written as

$$-\frac{1}{\eta\tan\overline{\theta}}=N.\tag{23}$$

It is noted that Eq. (23) is relatively simple, and the above-mentioned proof about universality is transformed into whether Eq. (21) can be simplified to Eq. (23). Expanding $\widetilde{\delta\theta}^{\text{inh}}(1)-\widetilde{\delta\theta}^{\text{inh}}(0)$ in the right term of Eq. (21) into a second-order form,

$$\widetilde{\delta\theta}^{\text{inh}}(1)-\widetilde{\delta\theta}^{\text{inh}}(0)=\left.\frac{\partial\widetilde{\delta\theta}^{\text{inh}}}{\partial X}\right|_{X=0}+\frac{1}{2}\left.\frac{\partial^2\widetilde{\delta\theta}^{\text{inh}}}{\partial X^2}\right|_{X=0}.\tag{24}$$

Thus, the numerator of the right term in Eq. (21) becomes

$$\begin{aligned}&(K_1-K_2)\widetilde{\delta\theta}^{\text{inh}}(0)-K_2\left[\widetilde{\delta\theta}^{\text{inh}}(1)-\widetilde{\delta\theta}^{\text{inh}}(0)\right]\\&=(K_1-K_2)\widetilde{\delta\theta}^{\text{inh}}(0)-K_2\left.\frac{\partial\widetilde{\delta\theta}^{\text{inh}}}{\partial X}\right|_{X=0}-\frac{K_2}{2}\left.\frac{\partial^2\widetilde{\delta\theta}^{\text{inh}}}{\partial X^2}\right|_{X=0}.\end{aligned}\tag{25}$$

Since $C_1\widetilde{\delta\theta}^{\text{inh}}(X)$ is the general solution of Eq. (7), $\widetilde{\delta\theta}^{\text{inh}}(X)$ needs to satisfy the homogeneous form of Eq. (7), namely,

$$\frac{K_2}{2}\left.\frac{\partial^2\widetilde{\delta\theta}^{\text{inh}}}{\partial X^2}\right|_{X=i}=(K_1-K_2)\widetilde{\delta\theta}^{\text{inh}}(i),\tag{26}$$

For the second-order partial derivative term in Eq. (25), by setting $i=0$ in Eq. (26), it can be expressed as

$$\frac{K_2}{2}\left.\frac{\partial^2\widetilde{\delta\theta}^{\text{inh}}}{\partial X^2}\right|_{X=0}=(K_1-K_2)\widetilde{\delta\theta}^{\text{inh}}(0).\tag{27}$$

For the first-order partial derivative term in Eq. (25), by summing Eq. (26) from $i=1$ to $i=N$ (see Appendix C for details), it can be expressed as

$$K_2 \frac{\partial \widetilde{\delta\theta}^{\mathrm{inh}}}{\partial X}\bigg|_{X=0} = -(K_1 - K_2)\sum_{i=1}^{N} \widetilde{\delta\theta}^{\mathrm{inh}}(i)\,. \tag{28}$$

Substituting Eqs. (27) and (28) into Eq. (25), Eq. (25) is simplified to

$$(K_1 - K_2)\widetilde{\delta\theta}^{\mathrm{inh}}(0) - K_2\left[\widetilde{\delta\theta}^{\mathrm{inh}}(1) - \widetilde{\delta\theta}^{\mathrm{inh}}(0)\right] = (K_1 - K_2)\sum_{i=1}^{N} \widetilde{\delta\theta}^{\mathrm{inh}}(i)\,. \tag{29}$$

Replacing the numerator of the right term in Eq. (21) by using Eq. (29), Eq. (21) is simplified to

$$-\frac{1}{\eta \tan\overline{\theta}} = \frac{\sum_{i=1}^{N} \widetilde{\delta\theta}^{\mathrm{inh}}(i)}{\widetilde{\delta\theta}^{\mathrm{inh}}(0)}\,. \tag{30}$$

Finally, all $\widetilde{\delta\theta}^{\mathrm{inh}}(i)$ $(i \in [1, N])$ in Eq. (30) are approximated as $\widetilde{\delta\theta}^{\mathrm{inh}}(0)$, and Eq. (30) is transformed into

$$-\frac{1}{\eta \tan\overline{\theta}} = \frac{\sum_{i=1}^{N} \widetilde{\delta\theta}^{\mathrm{inh}}(0) - \sum_{i=1}^{N}\left[\widetilde{\delta\theta}^{\mathrm{inh}}(0) - \widetilde{\delta\theta}^{\mathrm{inh}}(i)\right]}{\widetilde{\delta\theta}^{\mathrm{inh}}(0)} = N - \sum_{i=1}^{N}\left[1 - \frac{\widetilde{\delta\theta}^{\mathrm{inh}}(i)}{\widetilde{\delta\theta}^{\mathrm{inh}}(0)}\right]. \tag{31}$$

By comparing Eq. (31) and Eq. (23), it is indicated that Eq. (21) can be simplified to Eq. (23), but there is an error $\sum_{i=1}^{N}\left[1 - \widetilde{\delta\theta}^{\mathrm{inh}}(i) / \widetilde{\delta\theta}^{\mathrm{inh}}(0)\right]$. This error can be regarded as the truncation error generated by the zero-order Taylor expansion of $\widetilde{\delta\theta}^{\mathrm{inh}}(i)$ at $i = 0$, and this error is denoted as $R$. The value of $R$ is obtained by calculating its defining formula on curve $N + \left(\eta \tan\overline{\theta}\right)^{-1} = 0$ (the truncated form of the Eq. (31)), and when $R \ll N$ holds, both $\Delta L_e^{\mathrm{inh}}$ and $\Delta L_e^{\mathrm{h}}$ always tend to 0 simultaneously. In Appendix C, we demonstrate that $R \ll N$ always holds among the characteristic decay factors permitted by the anomalous size effect. So far, it has been proved that $\Delta L_e^{\mathrm{inh}}$ and $\Delta L_e^{\mathrm{h}}$ tend to 0 at the same time is universal. With the completion of this proof, the inference that the oscillating $P$-values is caused by $\Delta L_e^{\mathrm{inh}}$ and $\Delta L_e^{\mathrm{h}}$ tending to 0 simultaneously has been confirmed theoretically.

In the above proof, Eq. (23) gives the most concise relationship between the transformation size and the initial configuration ($N_P{}' = -\left(\eta \tan\overline{\theta}\right)^{-1} + 1$), which is noted by hollow circles in Fig. 4(d). The good agreement between the circle and the solid line confirms the correctness of the above theoretical analysis. In the experiment, by adjusting the original length of the spring, the initial configuration is adjusted to be consistent with the ordinate of the hollow circle. The stiffnesses at each size are measured under this initial configuration, and the size corresponding to the maximum stiffnesses are shown by the light pink dots in Fig. 4(d).

Eq. (23) also reveals another unusual property of bistable chains, that is, the transition size $N_P'$ is only dependent on the initial configuration and the shim length, neither the stiffness parameters of the shim nor the stiffness of the spring can change the value of $N_P'$. This anomalous effect is significantly different from the anomalous effects in the monostable chain, where the stiffness parameters of the flexible hinge influence the value of $N_P'$ (Coulais et al., 2018). The $N_P'$ of a bistable chain is related to only two physical quantities that describe the geometric relationship, which makes the design and customization of anomalous size effects easier. In addition, notably, Eq. (23) indicates that longer shims reduce $N_P'$. This may seem to contradict the relationship between $N_P'$ and hinge length in the monostable chain, but in fact, both are not in conflict. $\eta$ in Eq. (23) comes from $\eta l \sin\overline{\theta}$ in Eq. (2), and $\eta$ in Eq. (2) plays the role of describing the geometric relationship of the shim. The longer the shim, the greater the end-to-end displacement caused by $C_\theta$. $\Delta L_e^{\mathrm{h}}$ reaches 0 at the shorter nodes, resulting in the left shift of $N_P'$. However, in the monostable chain, $l_h$ plays the role of describing the equivalent stiffness of the flexible hinges. If $l_h$ becomes larger, intuitively it leads to a decrease in $k_\theta$ (more specifically, it leads to a decrease in $k_\theta / k_s$), and an increase in $\lambda^*$ $\left(\lambda^* \sim \sqrt{k_s / k_\theta}\right)$ causes the right shift of $N_P'$. Thus, $l_h$ plays different roles in the two kinds of chains, making it exhibit opposite effects on $N_P'$.

In addition to giving the relation Eq. (23), the above proof also shows that the oscillating $P$-values corresponds to that the inhomogeneous and homogeneous deformation can simultaneously approach 0. Namely, $\Delta L_e$ ($\Delta L_e = \Delta L_e^{\mathrm{inh}} + \Delta L_e^{\mathrm{h}}$) can approach to 0. In the case where the compliance can tend to 0, the maximum stiffness must occur at the point where the compliance tends to 0. Therefore, the curve obtained by connecting the discontinuity points must separate the increasing and decreasing interval, which provides a theoretical explanation for the stiffness taking $P$-values oscillation as the transition criticality under $\overline{\theta} = -\theta_s$. In physical implications, the negative $P$-values indicate that $\Delta L_e$ is the remaining deformation after inhomogeneous deformation and homogeneous deformation cancel each other out. The remaining deformation $\Delta L_e$ can be 0 in the above proof, which confirms that the cancellation mechanism does not through the opposite monotonicity to achieve non-monotonic stiffness, but through zero deformation. Specifically, the cancellation mechanism achieves the maximum stiffness of the structure by completely eliminating the deformation, then the stiffness is centered on the zero deformation point (oscillating *P*-values) and decreases to the two sides of the greater remaining deformation.

We show that above discussion about oscillating criticality can also provide an intuitive explanation for the constant criticality under $\overline{\theta} = \theta_s$. Based on the relation in Section 3 that the stiffnesses under the two steady states have the same maximum point (see Appendix C for details), the transition points for $\overline{\theta} = \theta_s$ and $\overline{\theta} = -\theta_s$ are denoted respectively as $(\overline{\theta}_s, N_P')$ and $(-\overline{\theta}_s, N_P')$. When $\Delta L_e^{\mathrm{inh}}(-\overline{\theta}_s, N_P') = 0$ and $\Delta L_e^{\mathrm{h}}(-\overline{\theta}_s, N_P') = 0$ are both satisfied, according to Eqs. (18) and (22),

$$\Delta L_e^{\mathrm{inh}}(\overline{\theta}_s, N_P') = \Delta L_e^{\mathrm{h}}(\overline{\theta}_s, N_P') = 2\Delta L_0. \tag{32}$$

Eq. (32) shows that when the structure size is $N_P'$, its inhomogeneous deformation is equal to its homogeneous deformation, namely, $P = 0.5$. This explains the constant criticality under $\overline{\theta} = \theta_s$, and this critical constant can only be 0.5, its value is not affected by any parameters, such as the parameters of the shim and the initial configuration. It can be seen that the cancellation mechanism simplifies the characterization of the maximum stiffness point under the competition mechanism. In other words, the stationary point problem corresponding to the competition mechanism is reduced order, and it is transformed into the zero point problem corresponding to the cancellation mechanism.

In this section, two kinds of criticalities correspond to two mechanisms for achieving the anomalous size effect. The one is the competition mechanism that regulates the non-monotonic variation of stiffness through opposite monotonicity. The other is the cancellation mechanism that forces the stiffness to decrease from the singular point to both sides through zero deformation. Among them, the cancellation mechanism is a new mechanism unique to bistable chains. For the anomalous size effect achieved by this cancellation mechanism. It should be noted that, since the $k_l$ of the shim is not infinite in reality, the tension deformation in Eq. (22) cannot be completely ignored. Therefore, the maximum stiffness under $\overline{\theta} = -\theta_s$ cannot reach infinity, such as $k_e(N')|_{N'=4}$ in bottom sub-figure of Fig. 2(d). However, although the stiffness has an upper bound, the relationship that $k_l$ is much greater than $k_s$ and $k_\theta l^{-2}$ is guaranteed by the flexible hinge, which makes the stiffness of the 4-node structure much larger than otherwise. This feature is obviously different from the flat stiffness curve under the competition mechanism, as shown in Fig. 2(d). In the following study in Section 5, the loading force is further increased. We will further discuss the difference between the two kinds of mechanisms under nonlinear large deformation.

**5. Robust anomalous size effect**

From the perspective of structural response, the anomalous size effect can be described as: when the excitation $F$ is fixed, the chain structure with more nodes has a smaller end-to-end displacement $\Delta L_e$. In the linear case discussed in the previous section, the displacement $\Delta L_e$ can be expressed as $\Delta L_e = (1 / k_e)F$. When a definite pair of $F$ is applied to both ends of the structure, the structure with greater stiffness $k_e$ will yield a smaller $\Delta L_e$. Therefore, in the previous section, when the stiffness of the large-size structure is larger than that of the small-size structure, the large-size structure has a smaller response, and the size effect of the structure is anomalous. In nonlinear large deformation, this scheme of size-displacement is continued to be used. The difference is that the secant stiffness ($k_{se}$) is used to characterize this size-deformation behavior under large deformation, because when $F$ is fixed, secant stiffness $k_{se}$ can also reflect the magnitude of $\Delta L_e$ ($\Delta L_e = (1 / k_{se})F$). Similarly, the anomalous size effect occurs when the secant stiffness of the large-size structure is greater than that of the small-size structure, because the large-size structure has a smaller response at this point.

In Fig. 5(a) and Fig. 5(b), we use the numerical solutions of the nonlinear model to study the evolution of the anomalous size-deformation behavior. All parameters in the numerical analysis are the same as those in Fig. 2. Fig. 5(a) shows the evolution of the anomalous size-deformation behavior based on the competition mechanism under tensile and compressive loads. In the case of tension, with the increase in $F$, the denser contours indicate that the variation trend of stiffness curve becomes steeper (the stiffness curve refers to the transversal of fixed $F$ and surface $k_{se}(N', F)$), the non-monotonic variation of stiffness is more obvious. In the case of compression, with the increase in $-F$, the contours become sparser, the stiffness curve becomes flatter, and as the curve flattens out, the gap between $k_{se}(N')|_{N'=2}$ and $k_{se}(N')|_{N'=4}$ gradually narrows. Taking the contour perpendicular to the vertical axis as the boundary, the anomalous size-deformation behavior disappears when $-F$ is larger than $0.12\,\mathrm{N}$. Therefore, the results of Fig. 5(a) show that this anomalous size-deformation behavior achieved by the competition mechanism is very sensitive to the type and magnitude of the load (or the nonlinear effect).

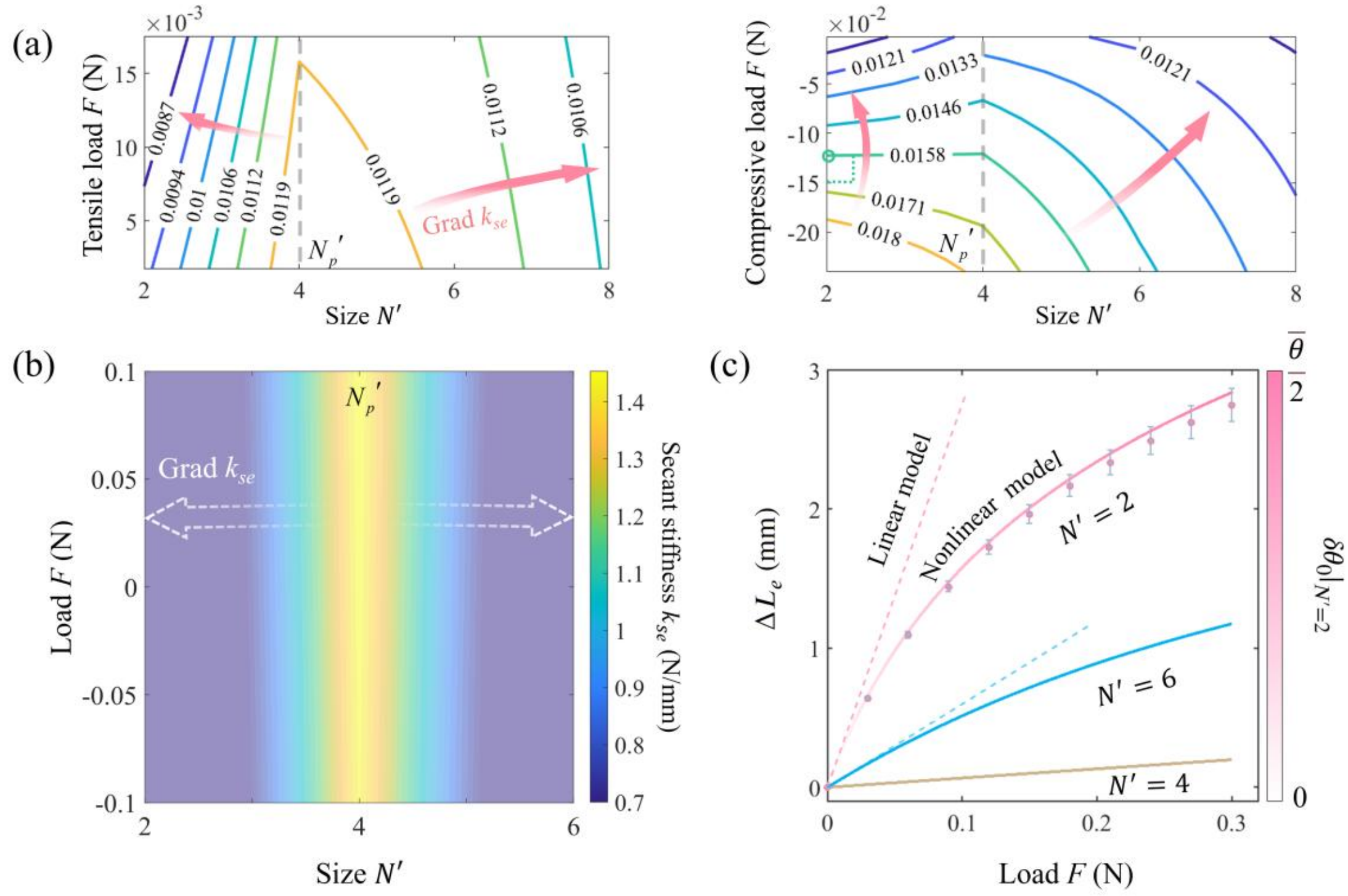

**Fig. 5.** (a) Contour plot of secant stiffness $k_{se}$ under the steady state $\overline{\theta} = \theta_s$ (the unit of contour line values is N/mm), where the positive and negative signs of the vertical axis $F$ respectively indicate that the structure is subjected to tension and pressure. The distribution of contour is affected by the type and magnitude of load $F$. In particular, the contour is perpendicular to the vertical axis when $F = -0.12$ N. (b) Contour plot of of secant stiffness under the steady state $\overline{\theta} = -\theta_s$, the distribution of contours is almost unaffected by the load $F$. (c) $F - \Delta L_e$ curves of $N' = \{2, 4, 6\}$, where the dotted lines are drawn based on the numerical results of the linear model in Section 3.1, and the solid curves are the numerical results of the nonlinear model. In the case of $N' = 2$, the depth of the curve color reflects the magnitude of the rotational deformation, and the allowable deformation amount of the 2-node structure (or the softest structure) limits the range of the load $F$. The hollow circle denotes the experimental result of the 2-node structure, and its error bars are calculated from three sets of experimental data.

Fig. 5(b) shows the evolution of the anomalous size-deformation behavior based on the cancellation mechanism. The results in the figure indicate that the anomalous behavior achieved by this new mechanism has strong robustness in nonlinear large deformation, which is shown by both the maximum stiffness and the maximum stiffness point are almost unaffected by the type and magnitude of loads. For the source of this robustness, it is analyzed in the next paragraph.

In the previous discussion, for the chain of $\overline{\theta} = -\theta_s$, the stiffness at $N' = 4$ is much greater than that in other cases, which makes the excitation required for the 4-node structure to enter the geometric nonlinear stage much greater than that of the structure with $N' \neq 4$. As an example, we plot the $F - \Delta L_e$ curve for the 2~6-node structure in Fig. 5(c). As can be seen from the figure, the 2-node and 6-node structures have produced significant nonlinear large deformation, even though the 4-node structures are still in the linear elastic stage. When the size effect is discussed under nonlinear large deformation, the selection of $F$ should ensure that the deformation of the softest structure is within an allowable range. Therefore, under such a $F$, the secant stiffness $k_{se}$ of the 4-node structure (or the hardest structure) hardly changes with the $F$ value. At the same time, the extremely large stiffness also means that the displacement response generated by the hardest structure is always the smallest, namely, the anomalous size-deformation behavior achieved by the cancellation mechanism is much more robust than that achieved by the competition mechanism.

## 6. Conclusions

In this work, anomalous size effects in bistable chains are systematically discussed. In a bistable chain, even in the case of linear small deformation, its translation and rotation are also coupled (more specifically, the angle is the differential of displacement), which gives the bistable chain an anomalous property — the end-to-end displacement difference caused by the decaying low-energy deformation is size-independent. This anomalous property not only makes the compliance of two steady states have the same change rate, but also establishes a connection between inhomogeneous deformation and homogeneous deformation, thus promoting the quantitative analysis of the anomalous size effect. In quantitative analysis, the proportion of inhomogeneous deformation faithfully reflects the monotonicity of stiffness, and the anomalous size effect has a constant or oscillating criticality under any initial configuration and material parameter. These two criticalities provide quantitative criteria for designing anomalous size effects in bistable chains. Even if they fail in the monostable case, Eq. (23) consistently gives the relationship between $N_P'$ and $\overline{\theta}$ in monostable and bistable chains assembled with low-torsion hinges. Because Eq. (23) diverges at $\overline{\theta} = 0$, which corresponds exactly to the result of the monostable chain after the tensile interactions are prevented (Coulais et al., 2018), namely, the stiffness does not decrease.

In addition, in the aspect of expanding the anomalous size effects, the cancellation mechanism makes it possible to combine this anomalous size-deformation behavior and nonlinearity robustly. But more importantly, for this class of anomalous size-deformation behavior that occurs at small sizes (such as the nanocrystalline metals (Schiøtz et al., 1998, Peng et al., 2022) and laminated beams (Wheel et al., 2015) soften as the size decreases), the cancellation mechanism provides a new consideration besides the competition mechanism.

**Acknowledgments**

This work was supported by National Natural Science Foundation of China (No. 12172183)，National Key Research and Development Program of China (No. 2023YFE0111000)，the Natural Science Foundation of Zhejiang Province (No. LZ24A020001), International Science and Technology Cooperation Project lauched by Science and Technology Bureau of Ningbo City, Zhejiang Province, China (No. 2023H011), One health Interdisciplinary Research Project (No. HY202206), Ningbo University.

## Appendix A. Governing equations

The elastic potential energy of a one-dimensional chain is

$$\begin{aligned} E = &\sum_{n=0}^{N-1} k_l\left\{u_{n+1} - u_n + \eta l\left[2\cos\overline{\theta} - \cos\left(\overline{\theta} + \delta\theta_n\right) - \cos\left(\overline{\theta} + \delta\theta_{n+1}\right)\right]\right\}^2 \\ &+ \sum_{n=0}^{N-1}\left\{k_s l^2\left[\sin\left(\overline{\theta} + \delta\theta_{n+1}\right) - \sin\left(\overline{\theta} + \delta\theta_n\right)\right]^2 + k_\theta\left(\delta\theta_n + \delta\theta_{n+1} + 2\overline{\theta}\right)^2\right\} \\ &+ \sum_{n=0}^{N}\frac{1}{2}a_n\left\{k_\theta\left(2\delta\theta_n + 2\overline{\theta}\right)^2 + k_t(\eta l)^2\left[2\cos\left(\overline{\theta} + \delta\theta_n\right) - \beta\right]^2\right\}, \end{aligned} \quad \text{(A.1)}$$

where, $\left[a_0, a_1, \cdots a_n \cdots, a_{N-1}, a_N\right] = [1/2, 1, \cdots 1 \cdots, 1, 1/2]$. The cases for linear small deformation ($\delta\theta_n \ll \overline{\theta}$) are considered. $\sin\left(\overline{\theta} + \delta\theta_n\right)$ and $\cos\left(\overline{\theta} + \delta\theta_n\right)$ in Eq. (A.1) are simplified as:

$$\sin\left(\overline{\theta} + \delta\theta_n\right) = \sin\overline{\theta} + \delta\theta_n\cos\overline{\theta} + o(\delta\theta_n), \quad \text{(A.2a)}$$

$$\cos\left(\overline{\theta} + \delta\theta_n\right) = \cos\overline{\theta} - \delta\theta_n\sin\overline{\theta} + o(\delta\theta_n). \quad \text{(A.2b)}$$

The simplified elastic potential energy is

$$\begin{aligned} E_{\text{lin}} = &\sum_{n=0}^{N-1} k_l\left[u_{n+1} - u_n + \eta l\sin\overline{\theta}\left(\delta\theta_n + \delta\theta_{n+1}\right)\right]^2 \\ &+ \sum_{n=0}^{N-1}\left\{k_s l^2\left(\delta\theta_{n+1} - \delta\theta_n\right)^2\cos^2\overline{\theta} + k_\theta\left(\delta\theta_n + \delta\theta_{n+1} + 2\overline{\theta}\right)^2\right\} \\ &+ \sum_{n=0}^{N}\frac{1}{2}a_n\left\{k_\theta\left(2\delta\theta_n + 2\overline{\theta}\right)^2 + k_t(\eta l)^2\left[2\cos\left(\overline{\theta} + \delta\theta_n\right) - \beta\right]^2\right\}. \end{aligned} \quad \text{(A.3)}$$

The relative displacement $\Delta L_e$ in Fig. 1(d) is connected with the displacement and rotation angle at both ends by the geometric constraint $c = 0$, where

$$c = \Delta L_e + u_N - u_0 - l\,(\delta\theta_0 \pm \delta\theta_N)\cos\overline{\theta}, \quad \text{(A.4)}$$

"+" and "−" correspond to the cases of even and odd nodes, respectively. Under the constraint $c = 0$, the Lagrange function is constructed as

$$L = E_{\text{lin}} + Fc, \quad \text{(A.5)}$$

where $F$ is the Lagrange multiplier, which corresponds to the pressure in Fig. 1(d). The equilibrium equation of the system is obtained when the objective function $E_{\text{lin}}$ reaches its minimum under geometric constraints. Let $\partial L/\partial u_n = 0$ to obtain the translational governing equation Eqs. (2), and let $\partial L/\partial\delta\theta_n = 0$ to obtain the rotational governing equation. In $\partial L/\partial\delta\theta_n$, the term with coefficients $k_l$ is simplified, namely,

$$k_l\left[u_{n+1}-u_n+\eta l\sin\overline{\theta}\left(\delta\theta_n+\delta\theta_{n+1}\right)\right]=-\frac{F}{2}. \tag{A.6}$$

Eq. (A.6) is obtained by the force balance of each cross in the horizontal direction. The term with coefficients $k_t$ is simplified as follows,

$$\begin{aligned}
&-k_t(\eta l)^2\left[2\cos\left(\overline{\theta}+\delta\theta_n\right)-\beta\right]\sin\left(\overline{\theta}+\delta\theta_n\right)\\
&=-k_t(\eta l)^2\left[2\cos\left(\overline{\theta}+\delta\theta_n\right)-\beta\right]\sin\left(\overline{\theta}+\delta\theta_n\right)+6k_\theta\overline{\theta}-6k_\theta\overline{\theta}\\
&=-k_t(\eta l)^2\left\{\sin\left(2\overline{\theta}+2\delta\theta_n\right)-\sin\left(2\overline{\theta}\right)+\left[\sin\overline{\theta}-\sin\left(\overline{\theta}+\delta\theta_n\right)\right]\beta\right\}-6k_\theta\overline{\theta}\\
&=-k_t(\eta l)^2\left[2\cos\left(2\overline{\theta}\right)-\beta\cos\overline{\theta}\right]\delta\theta_n-6k_\theta\overline{\theta}+o(\delta\theta_n),
\end{aligned} \tag{A.7}$$

where, the transformation of $6k_\theta\overline{\theta}$ in the second step is obtained by the initial equilibrium equation $6k_\theta\overline{\theta}=k_t(\eta l)^2\left[\sin\left(2\overline{\theta}\right)-\beta\sin\overline{\theta}\,\right]$. After being simplified by Eq. (A.6) and Eq. (A.7), $\partial L/\partial\delta\theta_n=0$ finally gives the governing equation Eqs. (3). Eqs. (2) and (3) are written in matrix form,

$$\begin{bmatrix}\mathbf{F}\\ \mathbf{M}\end{bmatrix}=\begin{bmatrix}\mathbf{K_t} & \mathbf{C}\\ \mathbf{O} & \mathbf{K_r}\end{bmatrix}\begin{bmatrix}\mathbf{U}\\ \boldsymbol{\delta\theta}\end{bmatrix}, \tag{A.8}$$

where, $\mathbf{F}=[-F/2,\quad 0,\quad \cdots\quad 0,\quad F/2]^{\mathrm{T}}$ is the horizontal external force matrix, $\mathbf{M}=\left[Fl\left(\eta\sin\overline{\theta}+\cos\overline{\theta}\right)/2,\quad F\eta l\sin\overline{\theta},\quad \cdots\quad F\eta l\sin\overline{\theta},\quad Fl\left(\eta\sin\overline{\theta}+\cos\overline{\theta}\right)/2\right]^{\mathrm{T}}$ is the moment matrix; $\mathbf{U}=[u_0,\quad \cdots\quad u_n,\quad \cdots\quad u_N]^{\mathrm{T}}$ and $\boldsymbol{\delta\theta}=[\delta\theta_0,\quad \cdots\quad \delta\theta_n,\quad \cdots\quad \delta\theta_N]^{\mathrm{T}}$ are the displacement and rotation angle matrices of each node. $\begin{bmatrix}\mathbf{K_t} & \mathbf{C}\\ \mathbf{O} & \mathbf{K_r}\end{bmatrix}$ is the block matrix form of the total stiffness matrix, and the submatrices $\mathbf{K_t}$ and $\mathbf{K_r}$ are the stiffness matrices of the translation and rotation components. Their specific forms are

$$\mathbf{K_r}=\begin{bmatrix}
k_l & -k_l & 0 & \cdots & \cdots & 0\\
-k_l & 2k_l & -k_l & 0 & \cdots & 0\\
0 & \ddots & \ddots & \ddots & \ddots & \vdots\\
\vdots & & \ddots & \ddots & \ddots & 0\\
\vdots & & & \ddots & 2k_l & -k_l\\
0 & \cdots & \cdots & \cdots & -k_l & k_l
\end{bmatrix}_{N'\times N'}, \tag{A.9}$$

and

$$\mathbf{K_t}=\begin{bmatrix}
K_1/2 & -K_2/2 & 0 & \cdots & \cdots & 0\\
-K_2/2 & K_1 & -K_2/2 & 0 & \cdots & 0\\
0 & \ddots & \ddots & \ddots & \ddots & \vdots\\
\vdots & & \ddots & \ddots & \ddots & 0\\
\vdots & & & \ddots & K_1 & -K_2/2\\
0 & \cdots & \cdots & \cdots & -K_2/2 & K_1/2
\end{bmatrix}_{N'\times N'}, \tag{A.10}$$

where, $K_1 = 4k_\theta + 2k_s l^2 \cos^2 \overline{\theta} - k_t(\eta l)^2 \left[2\cos\left(2\overline{\theta}\right) - \beta \cos \overline{\theta}\right]$ and $K_2 = 2\left(k_s l^2 \cos^2 \overline{\theta} - k_\theta\right)$. $\mathbf{C}$ is the coupling term generated by rotation in translational deformation. Correspondingly, the zero matrix $\mathbf{O}$ is the coupling term generated by translation in rotational deformation. The reason why this item is equal to 0 is that rotation and translation in rotational deformation are decoupled through Eq. (A.6). After obtaining the solution of Eq. (A.8), the end-to-end displacement is given by constraint $c = 0$, namely,

$$\Delta L_e = u_0 - u_N + l\,(\delta\theta_0 \pm \delta\theta_N)\cos\overline{\theta}. \tag{A.11}$$

**Appendix B. Supplement to the continuum model for even nodes**

In this Appendix, we supplement some details of the continuum model, which mainly includes three parts: (i) Solution for $\Delta L_0$, (ii) Solution for $\eta_e$ and (iii) Monotonicity of $\delta\theta_0^{\mathrm{inh}}(N')$.

(i) Solution for $\Delta L_0$

Substituting equation $\delta\theta(X) = \delta\theta^{\mathrm{inh}}(X) + C_\theta$ to Eqs. (3), the moment term $\left(F\eta l \sin\overline{\theta}\right)/2$ in Eqs. (3) can be eliminated by $C_\theta$, and the governing equation of inhomogeneous deformation is finally obtained, namely,

$$\begin{aligned}\left(Fl\cos\overline{\theta}\right)/2 = k_\theta\left(\delta\theta_1^{\mathrm{inh}} + 2\delta\theta_0^{\mathrm{inh}}\right) - k_s l^2\left(\delta\theta_1^{\mathrm{inh}} - \delta\theta_0^{\mathrm{inh}}\right)\cos^2\overline{\theta} \\ -(k_t/2)(\eta l)^2\left[2\cos\left(2\overline{\theta}\right) - \beta\cos\overline{\theta}\right]\delta\theta_0^{\mathrm{inh}},\end{aligned} \tag{B.1a}$$

$$\begin{aligned}0 = k_\theta\left(\delta\theta_{n+1}^{\mathrm{inh}} + \delta\theta_{n-1}^{\mathrm{inh}} + 4\delta\theta_n^{\mathrm{inh}}\right) - k_s l^2\left(\delta\theta_{n+1}^{\mathrm{inh}} + \delta\theta_{n-1}^{\mathrm{inh}} - 2\delta\theta_n^{\mathrm{inh}}\right)\cos^2\overline{\theta} \\ -k_t(\eta l)^2\left[2\cos\left(2\overline{\theta}\right) - \beta\cos\overline{\theta}\right]\delta\theta_n^{\mathrm{inh}}, n \in [1, N-1]\end{aligned} \tag{B.1b}$$

$$\begin{aligned}\left(Fl\cos\overline{\theta}\right)/2 = k_\theta\left(\delta\theta_{N-1}^{\mathrm{inh}} + 2\delta\theta_N^{\mathrm{inh}}\right) - k_s l^2\left(\delta\theta_{N-1}^{\mathrm{inh}} - \delta\theta_N^{\mathrm{inh}}\right)\cos^2\overline{\theta} \\ -(k_t/2)(\eta l)^2\left[2\cos\left(2\overline{\theta}\right) - \beta\cos\overline{\theta}\right]\delta\theta_N^{\mathrm{inh}}.\end{aligned} \tag{B.1c}$$

In Eqs. (B.1), the definition of $\delta\theta_n^{\mathrm{inh}}$ is $\delta\theta_n^{\mathrm{inh}} = \delta\theta^{\mathrm{inh}}(X)\,\big|_{X=n}$. Since $n \in [1, N-1]$, Eqs. (B.1) corresponds to $N'$ equations. Summing these $N'$ equations, the left term of the summation equation is $Fl\cos\overline{\theta}$. The right term of the summation equation is classified and discussed. For terms with coefficients $k_\theta$,

$$
\begin{aligned}
&\delta\theta_1^{\text{inh}} + 2\delta\theta_0^{\text{inh}} + \sum_{n=1}^{N-1}\left(\delta\theta_{n+1}^{\text{inh}} + \delta\theta_{n-1}^{\text{inh}} + 4\delta\theta_n^{\text{inh}}\right) + \delta\theta_{N-1}^{\text{inh}} + 2\delta\theta_N^{\text{inh}} \\
&= \left[\sum_{n=1}^{N-1}\left(\delta\theta_{n-1}^{\text{inh}}\right) + \delta\theta_{N-1}^{\text{inh}} + \delta\theta_1^{\text{inh}} + \sum_{n=1}^{N-1}\left(\delta\theta_{n+1}^{\text{inh}}\right)\right] + \left[2\delta\theta_0^{\text{inh}} + 4\sum_{n=1}^{N-1}\left(\delta\theta_n^{\text{inh}}\right) + 2\delta\theta_N^{\text{inh}}\right] \\
&= \sum_{n=0}^{N-1}\left(\delta\theta_n^{\text{inh}} + \delta\theta_{n+1}^{\text{inh}}\right) + 2\sum_{n=0}^{N-1}\left(\delta\theta_n^{\text{inh}} + \delta\theta_{n+1}^{\text{inh}}\right) \\
&= 3\sum_{0}^{N-1}\left(\delta\theta_n^{\text{inh}} + \delta\theta_{n+1}^{\text{inh}}\right). \qquad \text{(B.2)}
\end{aligned}
$$

For terms with coefficients $k_s$,

$$
\begin{aligned}
&\delta\theta_1^{\text{inh}} - \delta\theta_0^{\text{inh}} + \sum_{n=1}^{N-1}\left(\delta\theta_{n+1}^{\text{inh}} + \delta\theta_{n-1}^{\text{inh}} - 2\delta\theta_n^{\text{inh}}\right) + \delta\theta_{N-1}^{\text{inh}} - \delta\theta_N^{\text{inh}} \\
&= \delta\theta_1^{\text{inh}} - \delta\theta_0^{\text{inh}} + \sum_{n=1}^{N-1}\left[\left(\delta\theta_{n+1}^{\text{inh}} - \delta\theta_n^{\text{inh}}\right) - \left(\delta\theta_n^{\text{inh}} - \delta\theta_{n-1}^{\text{inh}}\right)\right] - \left(\delta\theta_N^{\text{inh}} - \delta\theta_{N-1}^{\text{inh}}\right) \qquad \text{(B.3)} \\
&= 0.
\end{aligned}
$$

For terms with coefficients $k_t$,

$$
\frac{1}{2}\delta\theta_0^{\text{inh}} + \sum_{n=1}^{N-1}\left(\delta\theta_n^{\text{inh}}\right) + \frac{1}{2}\delta\theta_N^{\text{inh}} = \frac{1}{2}\sum_{0}^{N-1}\left(\delta\theta_n^{\text{inh}} + \delta\theta_{n+1}^{\text{inh}}\right). \qquad \text{(B.4)}
$$

Combining Eqs. (B.2), (B.3) and (B.4), the summation equation is

$$
Fl\cos\overline{\theta} = \left\{3k_\theta - \frac{k_t(\eta l)^2}{2}\left[2\cos\left(2\overline{\theta}\right) - \beta\cos\overline{\theta}\right]\right\}\sum_{0}^{N-1}\left(\delta\theta_n^{\text{inh}} + \delta\theta_{n+1}^{\text{inh}}\right), \qquad \text{(B.5)}
$$

$\sum_0^{N-1}\left(\delta\theta_n^{\text{inh}} + \delta\theta_{n+1}^{\text{inh}}\right)$ is expressed as

$$
\sum_{0}^{N-1}\left(\delta\theta_n^{\text{inh}} + \delta\theta_{n+1}^{\text{inh}}\right) = \frac{2Fl\cos\overline{\theta}}{6k_\theta - k_t(\eta l)^2\left[2\cos\left(2\overline{\theta}\right) - \beta\cos\overline{\theta}\right]}. \qquad \text{(B.6)}
$$

Substituting Eq. (B.6) into $\Delta L_0 = \left(\eta l\sin\overline{\theta}\right)\sum_0^{N-1}\left(\delta\theta_n^{\text{inh}} + \delta\theta_{n+1}^{\text{inh}}\right)$,

$$
\Delta L_0 = \frac{F\eta l^2\sin\left(2\overline{\theta}\right)}{6k_\theta - k_t(\eta l)^2\left[2\cos\left(2\overline{\theta}\right) - \beta\cos\overline{\theta}\right]} = \frac{F\eta l^2\sin\left(2\overline{\theta}\right)}{K_1 - K_2}. \qquad \text{(B.7)}
$$

Combining Eq. (B.7) and the expression of $C_\theta$ (Eq. (10)), $\Delta L_0$ can be expressed simply as $2lC_\theta\cos\overline{\theta}$.

(ii) Solution for $\eta_e$

Substituting the angle solution Eq. (9) and the displacement solution Eq. (12)

into Eq. (A.11) yields

$$\Delta L_e = u_0^{\text{inh}} - u_N^{\text{inh}} + u_0^{\text{h}} - u_N^{\text{h}} + \left(\delta\theta_0^{\text{inh}} + \delta\theta_N^{\text{inh}} + 2C_\theta\right) l\cos\overline{\theta}, \tag{B.8}$$

where, $u_n^{\text{inh}} = u^{\text{inh}}(X)\big|_{X=n}$, $u_n^{\text{h}} = u^{\text{h}}(X)\big|_{X=n}$ and $\delta\theta_n^{\text{inh}} = \delta\theta^{\text{inh}}(X)\big|_{X=n}$. According to Eq. (B.8), the expressions for inhomogeneous deformation (denoted as $\Delta L_e^{\text{inh}}$) and homogeneous deformation (denoted as $\Delta L_e^{\text{h}}$) in Fig. 2(f) can be written as:

$$\Delta L_e^{\text{inh}} = u_0^{\text{inh}} - u_N^{\text{inh}} + \left(\delta\theta_0^{\text{inh}} + \delta\theta_N^{\text{inh}}\right) l\cos\overline{\theta}, \tag{B.9a}$$

$$\Delta L_e^{\text{h}} = u_0^{\text{h}} - u_N^{\text{h}} + 2C_\theta l\cos\overline{\theta}. \tag{B.9b}$$

Substituting the relationships $u_0^{\text{inh}} - u_N^{\text{inh}} = \Delta L_0$, $u_0^{\text{h}} - u_N^{\text{h}} = D_1 N$ and $\delta\theta_0^{\text{inh}} = \delta\theta_N^{\text{inh}}$ into Eqs. (B.9) yields

$$\Delta L_e^{\text{inh}} = \Delta L_0 + 2l\delta\theta_0^{\text{inh}}\cos\overline{\theta}, \tag{B.10a}$$

$$\Delta L_e^{\text{h}} = D_1 N + 2lC_\theta\cos\overline{\theta}. \tag{B.10b}$$

$\Delta L_e$ ($\Delta L_e = \Delta L_e^{\text{inh}} + \Delta L_e^{\text{h}}$) is finally represented as

$$\Delta L_e = 2\Delta L_0 + 2l\delta\theta_0^{\text{inh}} + D_1 N. \tag{B.11}$$

According to the expression $\eta_e = \Delta L_e / F$ for compliance,

$$\eta_e\left(\overline{\theta}, N'\right) = \frac{1}{F}\left[2\Delta L_0\left(\overline{\theta}\right) + 2l\delta\theta_0^{\text{inh}}\left(\overline{\theta}, N'\right) + D_1\left(\overline{\theta}\right)N\right]. \tag{B. 12}$$

For $\Delta L_0\left(\overline{\theta}\right)$ in Eq. (B.12), according to its expression Eq. (B.7), it has the symmetry $\Delta L_0\left(\overline{\theta}\right) = -\Delta L_0\left(-\overline{\theta}\right)$.

For $\delta\theta_0^{\text{inh}}\left(\overline{\theta}, N'\right)$ in Eq. (B.12),

$$\delta\theta_0^{\text{inh}}\left(\overline{\theta}, N'\right) = \frac{Fl\cos\overline{\theta}\left(1 + e^{N/\lambda^*}\right)}{K_1\left(1 + e^{N/\lambda^*}\right) - K_2\left[e^{1/\lambda^*} + e^{(N-1)/\lambda^*}\right]}. \tag{B.13}$$

Since $\lambda^*\left(\overline{\theta}\right) = \lambda^*\left(-\overline{\theta}\right)$, $K_1\left(\overline{\theta}\right) = K_1\left(-\overline{\theta}\right)$ and $K_2\left(\overline{\theta}\right) = K_2\left(-\overline{\theta}\right)$, $\delta\theta_0^{\text{inh}}\left(\overline{\theta}\right)$ has the symmetry $\delta\theta_0^{\text{inh}}\left(\overline{\theta}\right) = \delta\theta_0^{\text{inh}}\left(-\overline{\theta}\right)$.

For $D_1\left(\overline{\theta}\right)$ in Eq. (B.12),

$$D_1\left(\overline{\theta}\right) = \frac{2F\left(\eta l\sin\overline{\theta}\right)^2}{6k_\theta - k_t(\eta l)^2\left(2\cos\left(2\overline{\theta}\right) - \beta\cos\overline{\theta}\right)} + \frac{F}{2k_l}. \tag{B.14}$$

$D_1(\overline{\theta})$ has the symmetry $D_1(\overline{\theta}) = D_1(-\overline{\theta})$.

(iii) Monotonicity of $\delta\theta_0^{\mathrm{inh}}(N')$

According to Eq. (17), $\delta\theta_0^{\mathrm{inh}}$ can also be represented by

$$\delta\theta_0^{\mathrm{inh}} = \frac{Fl\cos\overline{\theta}}{K_1 - K_2\left[\frac{e^{1/\lambda^*} + e^{(N-1)/\lambda^*}}{1 + e^{N/\lambda^*}}\right]}, \tag{B.15}$$

$\delta\theta_0^{\mathrm{inh}}$ is always positive because $\delta\theta_0^{\mathrm{inh}}$ is the response under the positive moment $\left(Fl\cos\overline{\theta}\right)/2$. In the expression for $\delta\theta_0^{\mathrm{inh}}(N')$, the numerator is a positive constant independent of the size, which means that the monotonicity of $\delta\theta_0^{\mathrm{inh}}(N')$ can be determined by the monotonicity of the denominator.

A function $\varphi(N)$ is defined as

$$\varphi(N) = \frac{e^{1/\lambda^*} + e^{(N-1)/\lambda^*}}{1 + e^{N/\lambda^*}}. \tag{B.16}$$

In Eq. (B.15), $K_2 = 2\left(k_s l^2\cos^2\overline{\theta} - k_\theta\right) > 0$ is naturally satisfied for a low-torsion shim, so $\left(d\delta\theta_0^{\mathrm{inh}} / dN'\right)(d\varphi / dN') > 0$. Namely, the monotonicity of $\delta\theta_0^{\mathrm{inh}}(N')$ and $\varphi(N)$ is consistent. In order to analyze the monotonicity of $\varphi(N)$, $\varphi(N)$ is further expressed as

$$\varphi(N) = \frac{e^{1/\lambda^*} + e^{(N-1)/\lambda^*} + e^{-1/\lambda^*} - e^{-1/\lambda^*}}{1 + e^{N/\lambda^*}} = e^{-1/\lambda^*} + \frac{e^{1/\lambda^*} - e^{-1/\lambda^*}}{1 + e^{N/\lambda^*}}. \tag{B.17}$$

$\varphi(N)$ is monotonically decreasing, which makes $\delta\theta_0^{\mathrm{inh}}(N')$ also monotonically decreasing, thus $\delta\theta_0^{\mathrm{inh}}(N')$ reaches its maximum at the smallest even nodes $N' = 2$.

According to $\Delta L_e^{\mathrm{inh}} = 2l\left(C_\theta + \delta\theta_0^{\mathrm{inh}}\right)\cos\overline{\theta}$, the monotonically decreasing $\delta\theta_0^{\mathrm{inh}}(N')$ reveals a property of inhomogeneous deformation, namely, under the same force, the chain with a smaller size has a larger end-to-end displacement. This property is contrary to homogeneous deformation, and it is also contrary to the common perception that increasing size will soften the structure (Yang & Müller, 2021). This property is the fundamental reason for the increase of stiffness with size at small sizes.

Furthermore, it is noted that the monotonically decreasing property of $\varphi(N)$ is caused by the monotonically decreasing second term in Eq. (B.17). In order to clarify the cause of the monotonically decreasing $\varphi(N)$ from the physical perspective, Eq.

(B.17) is re-expressed as:

$$\varphi(N) = q^* + \Delta q(N), \tag{B.18}$$

where, $q^* = e^{-1/\lambda^*}$ and $\Delta q(N) = \left(e^{1/\lambda^*} - e^{-1/\lambda^*}\right) / \left(1 + e^{N/\lambda^*}\right)$. In terms of the physical meaning corresponding to the definition formula of $\varphi(N)$ (in Eq.(B.16), $\varphi(N) = \delta\theta_1^{\text{inh}} / \delta\theta_0^{\text{inh}}$), $\varphi(N)$ is the decay factor of the 0-th node in a finite structure. Eq. (B.18) expresses $\varphi(N)$ as the sum of $q^*$ and $\Delta q(N)$, where the size-independent $q^*$ is the characteristic decay factor of the infinitely long chain, and it is the decay factor corresponding to the deformation not generating any superposition effect. In contrast, the size-dependent $\Delta q(N)$ term can be understood as an additional term resulting from the superposition effect. Specifically, the connection between $\Delta q(N)$ and the superposition effect is as follows: $\Delta q(N)$ is constantly greater than 0, which reflects the reducing role of the superposition effect on the characteristic decay rate. In addition, $\Delta q(N)$ quantifies this reduction role. $\Delta q(N)$ decreases as the superposition effect weakens (the size $N'$ increases), and when the superposition effect disappears ($N' \to \infty$), $\Delta q(N) = 0$. In summary, the reason why the decay factor $\varphi(N)$ decreases with the increase of size is attributed to the weakening of the superposition effect under the large size, which re-examines the monotonicity of $\varphi(N)$ at the physical level. According to $\varphi(N)$ determining the monotonicity of $\delta\theta_0^{\text{inh}}$, the underlying physical mechanism of the anomalous size effect has been revealed, that is, the greater deformation caused by the stronger superposition effect in small-sized structures leads to structural softening.

The deformation distribution on the left can be expressed as

$$\delta\theta^{\text{inh}}(X) = C_1 e^{(N-X)/\lambda^*} + C_1 e^{X/\lambda^*}, \text{ for } X \to 0. \tag{B.19}$$

The deformation $C_1 e^{(N-X)/\lambda^*}$ is the local deformation on the left side, which corresponds to the decay deformation with a decay factor of $q^*$. The deformation $C_1 e^{X/\lambda^*}$ is the superposition of the right local deformation to the left local deformation. This is because its existence adds an additional term to the decay factor of each node in the finite structure. For a long enough chain, the local deformations at both ends almost do not produce superposition, and the deformation near the left end can be expressed as

$$\delta\theta^{\text{inh}}(X) = C_1 e^{(N-X)/\lambda^*}, \text{for } X \to 0. \tag{B.20}$$

Substituting Eq. (B.20) into Eq. (B.1a) yields

$$\delta\theta_0^{\mathrm{inh}} = \frac{Fl\cos\overline{\theta}}{K_1 - K_2 e^{-1/\lambda^*}}, \tag{B.21}$$

denoting the right term of Eq. (B.21) as $\delta\theta_0^{\mathrm{inh}*}$, $\delta\theta_0^{\mathrm{inh}*} = Fl\cos\overline{\theta} \,/\, \left(K_1 - K_2 e^{-1/\lambda^*}\right)$ is the characteristic deformation of the 0-th node in the semi-infinite chain. Since $\delta\theta_0^{\mathrm{inh}}(N')$ is monotonically decreasing and has a lower bound 0, $\delta\theta_0^{\mathrm{inh}*}$ is the limit of $\delta\theta_0^{\mathrm{inh}}(N')$, namely,

$$\lim_{N'\to\infty} \delta\theta_0^{\mathrm{inh}}(N') = \delta\theta_0^{\mathrm{inh}*}. \tag{B.22}$$

Eq. (B.22) shows that as the size increases, $\delta\theta_0^{\mathrm{inh}}(N')$ converges to a size-independent characteristic quantity $\delta\theta_0^{\mathrm{inh}*}$, which means inhomogeneous deformation can only lead to an increase of stiffness, the decreasing behavior of stiffness after the size reaches $N_P'$ is completely caused by homogeneous deformation.

Meanwhile, Eq. (B.22) also reveals the underlying logic that $\lambda^*$ can be used as the characteristic quantity of the anomalous size effects. In the process of size increase, the convergence of $\delta\theta_0^{\mathrm{inh}}(N')$ to $\delta\theta_0^{\mathrm{inh}*}$ actually corresponds to the convergence of the decay length $\lambda$ of the finite structure to the characteristic decay length $\lambda^*$, because both of these convergence can be equivalent to the approximation of Eq. (B.19) to Eq. (B.20). Therefore, we can estimate stiffness monotonicity based on the difference between $\lambda$ and $\lambda^*$, but we cannot uniformly give the critical point at which stiffness monotonicity transitions by making $\lambda - \lambda^*$ be equal to a specific value.

## Appendix C. Supplementation to the mathematical proof in Section 4.2

First, we supplement the derivation details of Eq. (21). Substituting Eqs. (10) and (17) into $\delta\theta_0^{\mathrm{inh}} = -C_\theta$ yields

$$\frac{Fl\cos\overline{\theta}\left(1 + e^{N/\lambda^*}\right)}{K_1\left(1 + e^{N/\lambda^*}\right) - K_2\left[e^{1/\lambda^*} + e^{(N-1)/\lambda^*}\right]} = -\frac{F\eta l\sin\overline{\theta}}{K_1 - K_2}. \tag{C.1}$$

Eq. (C.1) is transformed into

$$-\frac{1}{\eta\tan\overline{\theta}} = \frac{K_1\left(1 + e^{N/\lambda^*}\right) - K_2\left[e^{1/\lambda^*} + e^{(N-1)/\lambda^*}\right]}{(K_1 - K_2)\left(1 + e^{N/\lambda^*}\right)}. \tag{C.2}$$

In order to provide a path for the proof in Section 4.2, taking the $\left[\left(e^{1/\lambda^*}+e^{(N-1)/\lambda^*}\right)-\left(1+e^{N/\lambda^*}\right)\right]$ term as the construction target, the numerator of the right term in Eq. (C.2) is constructed as:

$$\begin{aligned}&K_1\left(1+e^{N/\lambda^*}\right)-K_2\left[e^{1/\lambda^*}+e^{(N-1)/\lambda^*}\right]\\&=K_1\left(1+e^{N/\lambda^*}\right)-K_2\left[e^{1/\lambda^*}+e^{(N-1)/\lambda^*}-\left(1+e^{N/\lambda^*}\right)+\left(1+e^{N/\lambda^*}\right)\right]\\&=(K_1-K_2)\left(1+e^{N/\lambda^*}\right)-K_2\left[e^{1/\lambda^*}+e^{(N-1)/\lambda^*}-\left(1+e^{N/\lambda^*}\right)\right].\end{aligned}\tag{C.3}$$

Thus, Eq. (C.2) becomes

$$-\frac{1}{\eta\tan\overline{\theta}}=\frac{(K_1-K_2)\left(1+e^{N/\lambda^*}\right)-K_2\left[e^{1/\lambda^*}+e^{(N-1)/\lambda^*}-\left(1+e^{N/\lambda^*}\right)\right]}{(K_1-K_2)\left(1+e^{N/\lambda^*}\right)}.\tag{C.4}$$

Eq. (C.4) is Eq. (21) in the main text.

For Eq. (28) in the main text, its derivation is based on Eq. (26) in the main text. Eq. (26) copied as follows:

$$\frac{K_2}{2}\frac{\partial^2\widetilde{\delta\theta}^{\mathrm{inh}}}{\partial X^2}\bigg|_{X=i}=(K_1-K_2)\widetilde{\delta\theta}^{\mathrm{inh}}(i).\tag{C.5}$$

Summing Eq. (C.5) from $i=1$ to $i=N$ yields

$$\frac{K_2}{2}\sum_{i=1}^{N}\frac{\partial^2\widetilde{\delta\theta}^{\mathrm{inh}}}{\partial X^2}\bigg|_{X=i}=(K_1-K_2)\sum_{i=1}^{N}\widetilde{\delta\theta}^{\mathrm{inh}}(i)\,.\tag{C. 6}$$

The backward difference form of the second-order partial derivative in Eq. (C.6) is

$$\frac{\partial^2\widetilde{\delta\theta}^{\mathrm{inh}}}{\partial X^2}\bigg|_{X=i}=\frac{\partial\widetilde{\delta\theta}^{\mathrm{inh}}}{\partial X}\bigg|_{X=i}-\frac{\partial\widetilde{\delta\theta}^{\mathrm{inh}}}{\partial X}\bigg|_{X=i-1}.\tag{C.7}$$

Summing Eq. (C.7) from $i=1$ to $i=N$ leads to

$$\sum_{i=1}^{N}\frac{\partial^2\widetilde{\delta\theta}^{\mathrm{inh}}}{\partial X^2}\bigg|_{X=i}=\sum_{i=1}^{N}\left(\frac{\partial\widetilde{\delta\theta}^{\mathrm{inh}}}{\partial X}\bigg|_{X=i}-\frac{\partial\widetilde{\delta\theta}^{\mathrm{inh}}}{\partial X}\bigg|_{X=i-1}\right)=\frac{\partial\widetilde{\delta\theta}^{\mathrm{inh}}}{\partial X}\bigg|_{X=N}-\frac{\partial\widetilde{\delta\theta}^{\mathrm{inh}}}{\partial X}\bigg|_{X=0}.\tag{C.8}$$

Eq. (C.8) is actually an approximate form of the definite integral for the second-order partial derivative. Since $\partial_X\widetilde{\delta\theta}^{\mathrm{inh}}$ is odd symmetric, namely, $\partial_X\widetilde{\delta\theta}^{\mathrm{inh}}\big|_{X=0}=-\partial_X\widetilde{\delta\theta}^{\mathrm{inh}}\big|_{X=N}$, Eq. (C.8) is finally expressed as

$$\sum_{i=1}^{N}\frac{\partial^2\widetilde{\delta\theta}^{\mathrm{inh}}}{\partial X^2}\bigg|_{X=i}=-2\frac{\partial\widetilde{\delta\theta}^{\mathrm{inh}}}{\partial X}\bigg|_{X=0}.\tag{C.9}$$

Based on Eq. (C.9), replacing $\sum_{i=1}^{N}\partial_{XX}\widetilde{\delta\theta}^{\mathrm{inh}}\big|_{X=0}$ in Eq. (C.6) with $-2\partial_X\widetilde{\delta\theta}^{\mathrm{inh}}\big|_{X=0}$, Eq. (C.6) becomes

$$-K_2 \frac{\partial \widetilde{\delta\theta}^{\mathrm{inh}}}{\partial X}\bigg|_{X=0} = (K_1 - K_2) \sum_{i=1}^{N} \widetilde{\delta\theta}^{\mathrm{inh}}(i)\,. \tag{C.10}$$

Eq. (C.10) is Eq. (28) in the main text.

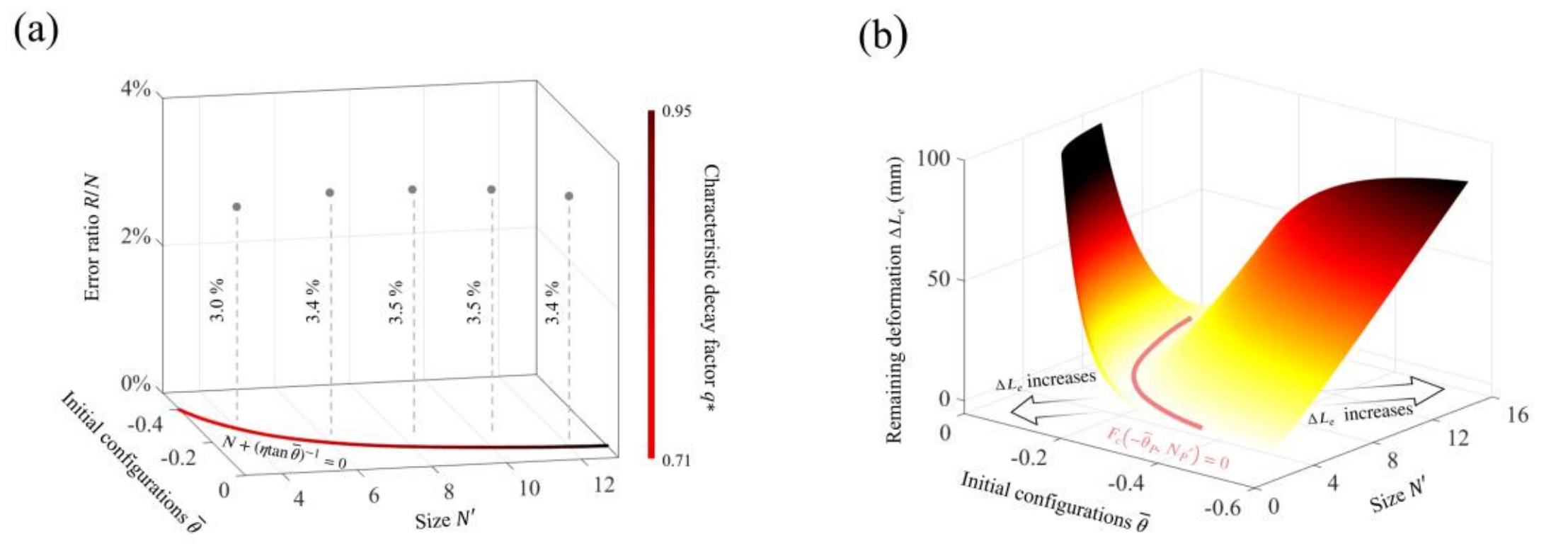

**Fig. C.** (a) The distribution figure of the error rate $R/N$ along curve $N + \left(\eta \tan \overline{\theta}\right)^{-1} = 0$, where the color of the curve represents the change of the characteristic decay factor. (b) The variation trend figure of the remaining deformation $\Delta L_e$. The $\Delta L_e$ reaches the minimum of 0 on curve $F_c\left(-\overline{\theta}_P, N_P{}'\right) = 0$ and increases to both sides, as indicated by the black arrows.

Next, we analyze the error $R$ mentioned in Section 4.2. The values of the error rate $R/N$ on curve $N + \left(\eta \tan \overline{\theta}\right)^{-1} = 0$ are shown in Fig. C(a). It can be seen that the error rate $R/N$ maintains a relatively smaller value on the curve; at the same time, it also has the following two features: (1) The values of $R/N$ at each size are close. (2) When the size is close to 2, the value of $R/N$ will be closer to 0. For the first feature, in the case of smaller size, the strong superposition effect significantly reduces the decay rate of inhomogeneous deformations, even if they have a larger characteristic decay factor (as shown by the color of the curve in Fig. C(a)); And for the case of larger size, although the superposition effect is weakened, the decay rate of the inhomogeneous deformation itself is very slow (the characteristicdecay factor is close to 1, as shown by the color of the curve in Fig. C(a)). Therefore, such a synergistic effect that $q^*$ must rise when the superposition effect weakens makes the values of $R/N$ close at each size. For the second feature, it is because $R$ ($R = \sum_{i=1}^{1}\left[\widetilde{\delta\theta}^{\mathrm{inh}}(i) - \widetilde{\delta\theta}^{\mathrm{inh}}(0)\right]$ in the 2-node structure is equal to 0, which makes the case where $N'$ is closer to 2 have a smaller error. The above analysis of errors can be extended to any large $q^*$ case, and the large $q^*$ case is what we mainly focus on, because for the case of small $q^*$, too fast decay rate will lead to the anomalous size effect that cannot be observed in discrete or real systems. Therefore, among the characteristic decay factors permitted by the anomalous size effect, $\sigma \ll N$ is considered to always hold.

Finally, we provide a supplementary explanation for the remaining deformation $\Delta L_e$. The remaining deformation $\Delta L_e$ is greater than or equal to 0, as shown in Fig. C(b). Such a $\Delta L_e$ can ensure that the relation $\partial_{N'}\left[\eta_e\left(\overline{\theta}, N'\right)\right] = \partial_{N'}\left[\eta_e\left(-\overline{\theta}, N'\right)\right]$ can always derive that the maximum stiffness points under the two steady states are consistent.